\documentclass[11pt]{article}
\usepackage{jcappub}
\usepackage{bm}
\usepackage{color}
\usepackage{graphicx}

\newcommand{\be}{\begin{equation}}
\newcommand{\ee}{\end{equation}}
\newcommand{\een}{\end{subequations}}
\newcommand{\ben}{\begin{subequations}}
\newcommand{\beq}{\begin{eqalignno}}
\newcommand{\eeq}{\end{eqalignno}}

\newcommand{\lsim}{\mathrel{\mathop{\kern 0pt \rlap
      {\raise.2ex\hbox{$<$}}}\lower.9ex\hbox{\kern-.190em $ \sim$}}}
\newcommand{\gsim}{\mathrel{\mathop{\kern 0pt
      \rlap{\raise.2ex\hbox{$>$}}}\lower.9ex\hbox{\kern-.190em $\sim$}}}

\DeclareMathOperator\erf{erf}

\title{Explaining DAMA with proton--philic spin--dependent inelastic
  Dark Matter (pSIDM): a frequentist analysis}

\author{Seong--Hyeong Kang}
\author{Stefano Scopel}
\author{Jong--Hyun Yoon}
\emailAdd{scopel@sogang.ac.kr}
\emailAdd{francis735@naver.com}
\emailAdd{pledge200@gmail.com}

\affiliation{Department of Physics, Sogang University, Seoul, South
  Korea}

\abstract{Proton--philic spin--dependent inelastic Dark Matter (pSIDM)
  is a Weakly Inelastic Massive Particle (WIMP) scenario that can
  explain the DAMA yearly modulation effect in compliance with the
  constraints from other direct detection Dark Matter (DM)
  searches. We obtain updated ranges for its parameters both in a
  halo--independent approach and adopting a truncated Maxwellian for
  the WIMP velocity distribution constructing approximate frequentist
  confidence intervals from an effective chi--square including, among
  others, the latest experimental constraints from XENON1T PANDAX--II,
  SuperCDMS and CDMSlite.  In the halo--independent analysis we have
  implemented the dependence on the WIMP velocity distribution through
  a step--wise parameterization of the halo function. Since in the
  pSIDM the WIMP incoming velocities required to explain the DAMA
  effect fall in a narrow range close to the escape velocity a limited
  number of steps was sufficient to determine the profile likelihood
  of the parameters. For the calculation of the profile likelihood we
  used a Markov Chain Montecarlo (MCMC) generator that required a
  large number of evaluations of the expected rates. To reduce
  computational time we introduced expected rates expressions
  rearranged in terms of differences of singled--valued integrated
  response functions suitable for a fast evaluation through tabulation
  and interpolation.  Our frequentist analysis confirms the present
  viability of the pSIDM scenario as a possible explanation of the
  DAMA effect. In terms of the WIMP mass $m_{\chi}$ and of the mass
  splitting $\delta$ we find the 1--sigma ranges $\mbox{12.5 GeV} \le
  m_{\chi}\le \mbox{15.7 GeV}$, $\mbox{22.1 keV} \le \delta \le
  \mbox{26.1 keV}$ for the halo--independent analysis and $\mbox{11.4
    GeV} \le m_{\chi}\le \mbox{13.6 GeV}$, $\mbox{24.4 keV} \le \delta
  \le \mbox{27.0 keV}$ for the Maxwellian case. We find that a full
  year of data taking of XENON1T should allow to start probing the
  pSIDM scenario.}

\begin{document}

\maketitle

\section{Introduction}
\label{sec:introduction}

A worldwide experimental effort is under way to detect Weakly
Interacting Massive Particles (WIMPs), which are considered the most
natural candidates to provide the Dark Matter (DM) in the halo of our
Galaxy, and the direct search for their recoils off nuclear targets
represents the most direct way to detect them
\cite{dama_last,xenon_1t,panda_2017,supercdms_2017,cdmslite2_2017,pico2l_2016,coupp}.
For quite a long time this search has been driven by theory, and
tailored on the search of the Supersymmetric neutralino, or on other
specific DM candidates provided by ultraviolet theoretical completions
of the Standard Model, believed to be observable in accelerator
physics. However the non-observation so far of new physics at the
Large Hadron Collider (LHC) has strongly prompted for the necessity to
go beyond this "to-down" approach in order to extend the search of
Dark Matter candidates to a wider range of properties through an
alternative "bottom-up" strategy not biased by theoretical prejudice.
Such a {\it model--independent} approach to WIMP direct searches must
address two issues: how to compare results from experiments using
different targets (i.e. which scaling law to adopt for the
WIMP--nucleus cross section) and which velocity distribution
determines the WIMP flux on Earth. Both these issues have been
addressed in the last few years: in particular, the WIMP--nucleus
cross section can be parameterized in terms of the most general
non--relativistic effective theory complying with Galilean symmetry
\cite{haxton1,haxton2} including modifications of the scattering
kinematics due to a possible inelasticity of the scattering process
\cite{inelastic}, while a halo--independent way to compare the results
of different experiments has been introduced by factorizing the
dependence of expected rates on the WIMP velocity distribution $f(v)$
in a single halo function $\eta(v)$\cite{factorization}.  In
particular, maximum-likelihood methods have been applied to determine
$\eta$ and the particle physics parameters, as well as statistical
methods to assess the compatibility of different experimental results
\cite{Feldstein_2014,Fox_2014,quantifying_feldstein,gondolo_out_of_the_bin,gelmini_assessing_compatibility,gelmini_convex_hulls,scopel_gondolo_unmodulated}.
In such analyses a likelihood function is introduced in terms of a
step--wise parameterization of the $\eta$ function (or, alternatively,
f(v) is expressed in terms of a sum of $\delta$ functions) that, under
some conditions, is proved to maximize the likelihood for a fixed
number of steps. However, while it is in principle straightforward to
find in this way the DM parameters that maximize the profile
likelihood function treating $\eta$ as a set of nuisance parameters,
as long as the presence of a DM signal in the existing data is not
established the result has little meaning since one would always
obtain a best--fit point even for highly incompatible data sets.

In the present paper we wish to apply such likelihood methods to a
specific WIMP scenario, where, instead, a very strong DM signal has
been established in the data and that has already been shown to allow
a full compatibility among different experiments: proton--philic
spin--dependent inelastic Dark Matter (pSIDM). Specifically, such
scenario was introduced in\cite{noi_idm_spin} to explain the DAMA
yearly modulation effect \cite{dama_last} in compliance with existing
constraints, and is summarized in Section \ref{sec:spin_idm_scenario}.
In particular, in the following we will use for the pSIDM a likelihood
function for which the theorems mentioned above on the number of steps
do not hold.  However in the pSIDM the WIMP incoming velocities
required to explain DAMA fall in a narrow range close to the escape
velocity so that, in practice, a limited number of steps in the $\eta$
parameterization will indeed be enough (this will be confirmed by
numerical inspection). In this way we will get updated ranges for the
pSIDM parameters $m_{\chi}$ (WIMP mass), $\delta$ (mass splitting) and
$r=c^n/c^p$ (neutron-to-proton coupling ratio) by constructing
approximate frequentist confidence intervals from an effective
chi--square including, among others, the latest experimental
constraints from XENON1T\cite{xenon_1t}, PANDAX-II\cite{panda_2017},
SuperCDMS\cite{supercdms_2017} and CDMSlite \cite{cdmslite2_2017}. Our
analysis will confirm the present viability of the pSIDM scenario as a
possible explanation of the DAMA effect. We will also integrate such
halo--independent result with a more standard analysis with the
velocity distributions f(v) given by a truncated Maxwellian.

For the calculation of the profile likelihood of the pSIDM parameters
we will use emcee\cite{emcee}, a Markov Chain Montecarlo (MCMC)
generator. The numerical procedure is straightforward but requires a
large number of evaluations of the expected rates for DAMA and the
other experiments included in the analysis (listed in Appendix
\ref{app:exp}). This would be time consuming if the relevant
experimental response functions were calculated at run time. For this
reason in Section \ref{sec:response_functions} we provide expected
rates expressions rearranged in terms of differences of singled--valued
functions that can be tabulated and interpolated.


\section{The pSIDM scenario and DAMA}
\label{sec:spin_idm_scenario}

In this Section we briefly summarize the features of the scenario
introduced in Ref.\cite{noi_idm_spin} (we refer the reader to such
paper for further details).

The most stringent bounds on an interpretation of the DAMA effect in
terms of WIMP--nuclei scatterings are obtained by detectors using
xenon (LUX\cite{xenon_1t,lux,lux_2015_reanalysis,lux_complete},
PANDA\cite{panda_2017,panda_run_8,panda_run_9}) and germanium
(CDMS\cite{supercdms_2017,cdms_ge,cdmslite2_2017,cdms_lite,super_cdms,cdms_2015})
whose spin is mostly originated by an unpaired neutron, as well as by
the KIMS experiment\cite{kims,kims_modulation,kims2} which uses $CsI$
and thus directly probes the contribution to the DAMA effect from WIMP
scatterings off iodine targets. If the WIMP mass is small enough to
assume that the DAMA signal is only due to WIMP scatterings off sodium
the KIMS constraint can be evaded. Moreover, both sodium and iodine in
DAMA have an unpaired proton, so that if the WIMP particle interacts
with ordinary matter predominantly via a spin--dependent coupling
which is suppressed for neutrons it can explain the DAMA effect in
compliance with the bounds from xenon and germanium detectors, whose
constraints are strongly
relaxed\cite{spin_n_suppression,spin_gelmini}. However this scenario
is constrained by droplet detectors (SIMPLE\cite{simple},
COUPP\cite{coupp}) and bubble chambers (PICASSO\cite{picasso},
PICO-2L\cite{pico2l,pico2l_2016},PICO-60\cite{pico60}) which all use
nuclear targets with an unpaired proton (in particular, they all
contain $^{19}F$, while SIMPLE contains also $^{35}Cl$ and $^{37}Cl$
and COUPP and PICO-60 use also $^{127}I$).  As a consequence, this
class of experiments rules out a DAMA explanation in terms of WIMP
elastic scatterings with a spin--dependent coupling to protons when
standard assumptions are made on the WIMP local density and velocity
distribution in our Galaxy\cite{spin_gelmini,pico2l}.

In Ref.\cite{noi_idm_spin} the alternative approach of Inelastic Dark
Matter (IDM) was proposed to reconcile DAMA to fluorine detectors. In
this class of models a DM particle $\chi_1$ of mass
$m_{\chi_1}=m_{\chi}$ interacts with atomic nuclei exclusively by
up--scattering to a second heavier state $\chi_2$ with mass
$m_{\chi_2}=m_{\chi}+\delta$. A peculiar feature of IDM is that there
is a minimal WIMP incoming speed in the lab frame matching the
kinematic threshold for inelastic upscatters and given by:

\begin{equation}
v_{min}^{*}=\sqrt{\frac{2\delta}{\mu_{\chi N}}},
\label{eq:vstar}
\end{equation}

\noindent with $\mu_{\chi N}$ the WIMP--nucleus reduced
mass. This quantity corresponds to the lower bound of the minimal
velocity $v_{min}$ (also defined in the lab frame) required to deposit
a given recoil energy $E_R$ in the detector:

\begin{equation}
v_{min}=\frac{1}{\sqrt{2 m_N E_R}}\left | \frac{m_NE_R}{\mu_{\chi N}}+\delta \right |,
\label{eq:vmin}
\end{equation}

\noindent with $m_N$ the nuclear mass. In particular, indicating with
$v_{min}^{*Na}$ and $v_{min}^{*F}$ the values of $v_{min}^*$ for
sodium and fluorine, and with $v_{cut}$ the result of the boost in the
lab rest frame of some maximal speed value beyond which the WIMP
velocity distribution $f(v)$ in the galactic rest frame vanishes
(typically $v_{cut}$ is identified with the WIMP escape velocity
$v_{esc}$), constraints from droplet detectors and bubble chambers can
be evaded when the WIMP mass $m_{\chi}$ and the mass gap $\delta$ are
chosen in such a way that the hierarchy:

\begin{equation}
v_{min}^{*Na}<v_{cut}^{lab}<v_{min}^{*F},
\label{eq:hierarchy}
\end{equation}

\noindent is achieved, since in such case WIMP scatterings off
fluorine turn kinematically impossible while those off sodium can
still serve as an explanation to the DAMA effect. Clearly, this
mechanism rests on the trivial observation that the velocity
$v_{min}^*$ for fluorine is larger than that for sodium.

As a consequence of what discussed above, in the pSIDM scenario the
WIMP--nucleus interaction must be fixed to a spin--dependent coupling,
which in the most simple case corresponds to

\begin{equation}
{\cal L}_{int}\ni
c^p\vec{S}_{\chi}\cdot \vec{S}_p+c^n\vec{S}_{\chi}\cdot \vec{S}_n,
\label{eq:spin_coupling}
\end{equation}
\noindent with ${\cal N}=p,n$, while the parameter $r=c^n/ c^p\ll1$
will be allowed to vary freely and its range will be determined by the
likelihood analysis of Section \ref{sec:results}.  

\section{Expected rates and response functions}
\label{sec:expected_rates}

\label{sec:response_functions}
The expected rate in a given visible energy bin $E_1^{\prime}\le
E^{\prime}\le E_2^{\prime}$ of a direct detection experiment is given
by:

\begin{eqnarray}
R_{[E_1^{\prime},E_2^{\prime}]}&=&MT\int_{E_1^{\prime}}^{E_2^{\prime}}\frac{dR}{d
  E^{\prime}}\, dE^{\prime} \label{eq:start}\\
 \frac{dR}{d E^{\prime}}&=&\sum_T \int_0^{\infty} \frac{dR_{\chi T}}{dE_{ee}}{\cal
   G}_T(E^{\prime},E_{ee})\epsilon(E^{\prime})\label{eq:start2}\,d E_{ee} \\
E_{ee}&=&q(E_R) E_R \label{eq:start3},
\end{eqnarray}

\noindent with $\epsilon(E^{\prime})\le 1$ the experimental
efficiency/acceptance. In the equations above $E_R$ is the recoil
energy deposited in the scattering process (indicated in keVnr), while
$E_{ee}$ (indicated in keVee) is the fraction of $E_R$ that goes into
the experimentally detected process (ionization, scintillation, heat)
and $q(E_R)$ is the quenching factor, ${\cal
  G_T}(E^{\prime},E_{ee}=q(E_R)E_R)$ is the probability that the
visible energy $E^{\prime}$ is detected when a WIMP has scattered off
an isotope T in the detector target with recoil energy $E_R$, $M$ is
the fiducial mass of the detector and $T$ the live--time of the data
taking. Moreover,

\begin{equation}
\frac{dR_{\chi T}}{dE_R}=N_T \int_{v_{min}}^{v_{esc}} \frac{\rho_{\chi}}{m_{\chi}}
v \frac{d\sigma_{\chi T}}{dE_R}\, f(v) dv,
\label{eq:dr_der}  
\end{equation}

\noindent where $N_T$ is the number of targets T per unit mass, while
$\rho_{\chi}$ and $m_{\chi}$ are the WIMP local density and mass. The
most general WIMP--nucleon interaction \cite{noi_eft_spin} (including
momentum and velocity dependence) can be parameterized by making use
of the interaction Hamiltonian which descends from non--relativistic
EFT\cite{haxton1,haxton2}:

\begin{eqnarray}
{\bf\mathcal{H}}&=& \sum_{\tau=0,1} \sum_{k=1}^{15}
c_k^{\tau} \mathcal{O}_{k} \, t^{\tau} ,
\label{eq:H}
\end{eqnarray}

\noindent where $t^0=1$, $t^1=\tau_3$ denote the the $2\times2$
identity and third Pauli matrix in isospin space, respectively, the
dimensional -2 isoscalar and isovector coupling constants $c^0_k$ and
$c^{1}_k$ are related to those to protons and neutrons $c^{p}_k$ and
$c^{n}_k$ by $c^{p}_k=(c^{0}_k+c^{1}_k)/2$ and
$c^{n}_k=(c^{0}_k-c^{1}_k)/2$ and the operators ${\cal O}_i$ are for
instance listed in Equations (12) and (13) of \cite{haxton2}. Using
the notation of \cite{haxton2}:

\begin{equation}
\frac{d\sigma_{\chi T}}{d E_R}=\frac{1}{10^6}\frac{2
  m_T}{4\pi}\frac{c^2}{v^2}\left [\frac{1}{2j_{\chi}+1}
  \frac{1}{2j_{T}+1}\sum_{spin}|{\cal M_T}|^2 \right ],
\label{eq:haxton_50}
\end{equation}

\noindent with:

\begin{equation}
  \frac{1}{2j_{\chi}+1} \frac{1}{2j_{T}+1}\sum_{spin}|{\cal M_T}|^2=
  \frac{4\pi}{2j_T+1}\sum_{\tau\tau^{\prime}}\sum_l R_l^{\tau\tau^{\prime}} W_{T,l}^{\tau\tau^{\prime}},
\label{eq:haxton_40}
\end{equation}

\noindent and, including velocity--dependent terms:

\begin{equation}
R_l^{\tau\tau^{\prime}}=R_{0,l}^{\tau\tau^{\prime}}+R_{1,l}^{\tau\tau^{\prime}}(v^2-v_{min}^2).
  \end{equation}

The factor $10^{-6}$ in front of Eq. (\ref{eq:haxton_50}) is to
express the differential cross section in GeV$^{-2}$/keV if the
$c_k^{\tau}$ couplings are expressed in GeV$^{-2}$. The WIMP response
functions $R_l^{\tau\tau^{\prime}}$ are provided in Eq.(38) of
\cite{haxton2} (but in the equation above the factor $q^2/m_n^2$ that
multiplies the last 5 terms in Eq. (40) of \cite{haxton2} have been
incorporated in the definitions of the $R_l^{\tau\tau^{\prime}}$'s)
and $l$=$M$,$\Sigma^{\prime\prime}$
,$\Sigma^{\prime}$,$\Phi^{\prime\prime}$,$\Phi^{\prime\prime}M$,
$\tilde{\Phi}^{\prime}$,$\Delta$, $\Delta\Sigma^{\prime}$ represent
one of the possible nuclear interaction types.  As far as the pSIDM
scenario is concerned, it can be implemented with any interaction
operator with an exclusively spin--dependent nuclear response function
(i.e. depending only on either the function
$W^{\tau\tau^{\prime}}_{\Sigma^{\prime}}$ or
$W^{\tau\tau^{\prime}}_{\Sigma^{\prime\prime}}$ in the notation of
\cite{noi_idm_spin}). Such operators, for instance summarized in Table
2 of Ref. \cite{noi_eft_spin}, correspond to $k$=4, 6, 7, 9, 10 and
14.  In particular, in the present analysis we only consider the
standard spin dependent case, which in Eq.(\ref{eq:H}) and the
notation of (\cite{haxton1,haxton2}) corresponds to $k$=4.  In the
case of a single coupling we factorize the {\it conventional }
WIMP--proton cross section:
\begin{equation}
\sigma_p=(c_4^p)^2\frac{\mu_{\chi{\it N}}^2}{\pi},
  \label{eq:conventional_sigma}
  \end{equation}
\noindent with $\mu_{\chi{\it N}}$ the WIMP--nucleon reduced
mass\footnote{The standard WIMP--proton cross section corresponds to
  3/16 $\sigma_p$.}.  Combining everything together and inverting the
order of the two integrals in $E_R$ and $v$\cite{gondolo_generalized},
for inelastic scattering one gets:

\begin{equation}
  \int_0^{\infty}d E_R\int_{v_{min}}^{\infty}d v\rightarrow
  \int_{v^*_{min}}^{\infty} d v \int_{E^{-}_{R}(v)}^{E^{+}_{R}(v)} d E_R,
\end{equation}

\noindent where the quantities $E_R^{\pm}(v,m_{\chi},\delta)$
represent the two values of $E_R$ which correspond to the same
$v_{min}$, i.e.:

\begin{equation}
  E_R^{\pm}(v_{min},m_{\chi},\delta)=\frac{\mu_{\chi N}^2}{m_N}\left [ v_{min}^2-\frac{(v_{min}^*)^2}{2} 
  \pm \sqrt{v_{min}^2-(v_{min}^*)^2}\right ],
\end{equation}

\begin{figure}
\begin{center}
  \includegraphics[width=0.8\columnwidth]{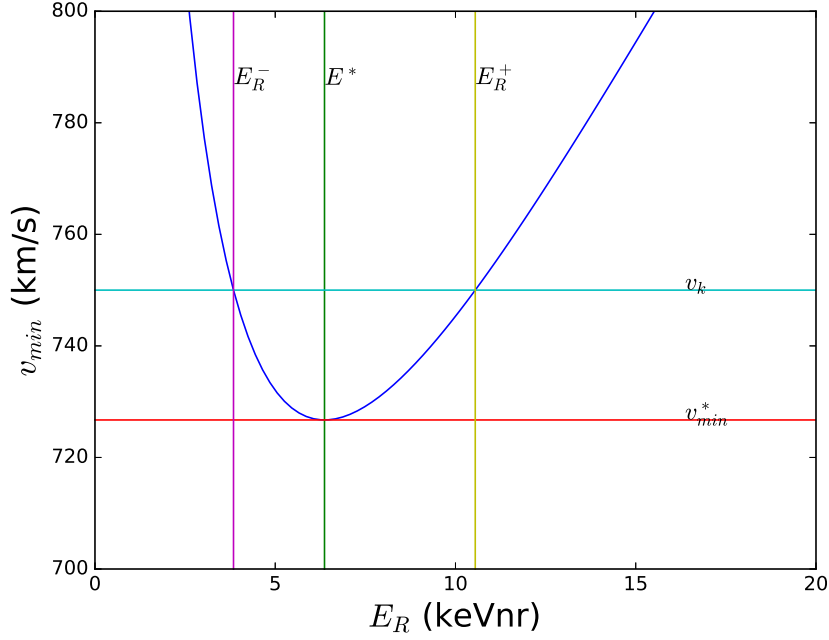}
\end{center}
\caption{Mapping between $E_R$ and $v_{min}$ in $^{23}$Na for
  $m_{\chi}$=10 GeV and $\delta$=20 keV.}
\label{fig:er_vmin}
\end{figure}

\noindent (a schematic view of the mapping between $E_R$ and $v_{min}$
for inelastic scattering is illustrated in Fig.\ref{fig:er_vmin}).
Then:

\begin{equation}
  R_{[E_1^{\prime},E_2^{\prime}]}=\frac{\rho_{\chi}}{m_{\chi}} \sigma_p
  \int_{v^*_{min}}^{\infty} d v \hat{{\cal H}}(v) f(v),
  \end{equation}

\noindent with:

\begin{eqnarray}
  \hat{{\cal H}}(v)&=&\sum_T N_T MT \frac{c^2}{v}\frac{m_T}{\mu_{\chi T}^2}\frac{2\pi}{10^6}
  \int_{E^{-}_{R}(v)}^{E^{+}_{R}(v)} d E_R \int_{E_1^{\prime}}^{E_2^{\prime}} dE^{\prime} \epsilon(E^{\prime}) {\cal G}_T[E^{\prime},q(E_R)E_R]\nonumber \times\\
  && \frac{1}{2j_T+1} \sum_{\tau\tau^{\prime}}\sum_l \left [\hat{R}_{0,l}^{\tau\tau^{\prime}}+\hat{R}_{1,l}^{\tau\tau^{\prime}}(v^2-v_{min}^2)\right ] W_l^{\tau\tau^{\prime}}=\\
&& \frac{c^2}{v} \int_{E^{-}_{R}(v)}^{E^{+}_{R}(v)} d E_R \left \{ \hat{{\cal R}}_0(E_R)+ \hat{{\cal R}}_1(E_R)(v^2-v_{min}(E_R)^2)\right \},
  \end{eqnarray}

\noindent and:

\begin{eqnarray}
&&\hat{{\cal R}}_{\{0,1\}}=\sum_T N_T MT \frac{m_T}{\mu_{\chi
    T}^2}\frac{2\pi}{10^6} \int_{E_1^{\prime}}^{E_2^{\prime}}
dE^{\prime} \epsilon(E^{\prime}) {\cal
  G}_T[E^{\prime},q(E_R)E_R]\frac{1}{2j_T+1}
\sum_{\tau\tau^{\prime}}\sum_l \hat{R}_{\{0,1\},l}^{\tau\tau^{\prime}}
W_l^{\tau\tau^{\prime}}\nonumber\\
&&=\sum_T\left [\hat{{\cal R}}_{\{0,1\}} \right ]_T,
\label{eq:response_function_final}
\end{eqnarray}

\noindent where $\hat{R}_{0,l}^{\tau\tau^{\prime}}\equiv
R_{0,l}^{\tau\tau^{\prime}}/(c^p_4)^2$. Setting\cite{gondolo_generalized}:
\begin{equation}
f(v)\equiv -v\frac{d}{dv}\eta(v),
  \end{equation}
\noindent and integrating by parts one gets:

\begin{equation}
  R_{[E_1^{\prime},E_2^{\prime}]}=\frac{\rho_{\chi}}{m_{\chi}} \sigma_p
  \int_0^{\infty} dv \hat{{\cal R}}(v)\eta(v)=\int_0^{\infty} dv {\cal
    R}(v)\tilde{\eta}(v), 
  \label{eq:int_r_eta}
\end{equation}

\noindent with: 

\begin{equation}
  \tilde{\eta}(v,t)=\frac{\rho_{\chi}}{m_{\chi}}\sigma \eta(v,t),\,\,\,\,
  \eta(v,t)=\int_{v}^{\infty}\frac{f(v,t)}{v}\,dv,
\label{eq:eta_tilde}  
\end{equation}

\noindent and where now the response function $\hat{{\cal R}}$ depends
on $\hat{{\cal H}}$ through:

\begin{equation}
  \hat{{\cal R}}(v)=\frac{d}{dv}\left [ v \hat{{\cal H}}(v) \right ].
  \label{eq:r_h}
  \end{equation}

\noindent In Eq.(\ref{eq:eta_tilde}) we have explicitly indicated that
the the generalized halo function $\tilde{\eta}$ depends, as the
velocity distribution $f(v)$, on time $t$, due to the rotation of the
Earth around the Sun.  In particular, the generalized halo function
$\tilde{\eta}(v)$ is common to all experiments, while the response
function ${\cal R}(v)$ depends on experimental inputs. Notice that in
Eq.(\ref{eq:eta_tilde}) all quantities depend on $v=|\vec{v}|$ because
present experiments have no directional sensitivity, so ${\cal
  R}(\vec{v})$=${\cal R}(v)$. This means that in (\ref{eq:eta_tilde})
one has $\tilde{\eta}(v)\equiv \int\, d\Omega\,v^2
\tilde{\eta}(\vec{v})$ and $f(v)\equiv \int\, d\Omega\,v^2
f(\vec{v})$, i.e. present experiments are not sensitive to the angular
dependence of the velocity distribution if the latter is expressed in
the detector's reference frame. Due to this latter property, that
ensures the validity of Eq.(\ref{eq:int_r_eta}) for any velocity
distribution, in the following we choose to express the rate in the
lab rest frame.

Given a time--dependent signal $S(t)\equiv
R_{[E_1^{\prime},E_2^{\prime}]}(t)$, present direct detection
experiments have either access to the time average of $S$, i.e.:

\begin{eqnarray}
S_0&\equiv& \frac{1}{T}\int_0^T S(t)dt=\int_{0}^{\infty} {\cal R}(v)
\tilde{\eta}_0(v)\,dv, \nonumber\\
\tilde{\eta}_0(v)&\equiv& \frac{1}{T}\int_0^T \tilde{\eta}(v,t)dt,
\label{eq:s0}
\end{eqnarray}  

\noindent or, as in the case of the DAMA experiment, to the yearly
modulation amplitude $S_1$, defined as the cosine transform of S:

\begin{eqnarray}
S_1&\equiv& \frac{2}{T}\int_0^T
\cos\left[\frac{2\pi}{T}(t-t_0)\right]S(t)dt=\int_{0}^{\infty} {\cal R}(v)
\tilde{\eta}_1(v)\,dv, \nonumber\\
\tilde{\eta}_1(v)&\equiv&
\frac{2}{T}\int_0^T \cos\left[\frac{2\pi}{T}(t-t_0)\right] \tilde{\eta}(v,t)dt,
\label{eq:s1}
\end{eqnarray}  

\noindent with $T$=1 year and $t_0$=2 June.  In the halo independent
method no assumptions are made on the velocity distribution $f$, so
that the two halo functions $\tilde{\eta}_0(v)$ and
$\tilde{\eta}_1(v)$ are subject to the very general conditions:

\begin{eqnarray}
&\tilde{\eta}_0(v_{2})\le\tilde{\eta}_0(v_{1})  & \mbox{if $v_{2}> v_{1}$},\nonumber\\ 
& |\tilde{\eta}_1|\le\tilde{\eta}_0  & \mbox{at the same $v$},\nonumber\\
& \tilde{\eta}_0(v \ge v_{esc})=0, & 
\label{eq:eta_conditions}
\end{eqnarray}

\noindent with $v_{esc}$ the galactic escape velocity expressed in the
lab rest frame. In particular, it has been recently shown that, for a
given set of annual modulation experimental data, if the effect is
totally ascribed to the time--dependent change of reference frame
between the lab and the Galaxy, even in a halo-independent approach it
is possible to improve the second constraint of
Eq.(\ref{eq:eta_conditions}), i.e. to get
$|\tilde{\eta}_1/\tilde{\eta}_0|<a$ with $a<$1 (namely, until now only
an analysis of the DAMA data restricted to velocity distributions
which are isotropic in the galactic rest frame is available, with
values of $a$ varying from $\simeq$ 0.14 and $\simeq$ 0.25 depending
on the WIMP mass \cite{gondolo_scopel}, although the same procedure
can in principle be applied to the non--isotropic case). However,
allowing for a possible variation of the WIMP local density $\rho$
with the Earth's position, the general range of $\tilde{\eta}_1$ given
in Eq.(\ref{eq:eta_conditions}) is in principle always saturated.

The continuous halo function $\tilde{\eta}(v)$ depends in principle on
an infinite number of parameters. However, for practical purposes a
possible approach to this problem is to parameterize
$\tilde{\eta}_0(v)$ with a step function sampled in a large--enough
number of velocity steps, i.e., to set:

\begin{equation}
  \tilde{\eta}_{0,1}(v)= \sum_{k=1}^{N}
  \tilde{\eta}_{0,1}^k\theta(v-v_{k-1})\theta(v_k-v),
\label{eq:eta01_piecewise}
\end{equation}  

or, equivalently:

\begin{eqnarray}
&&  \tilde{\eta}_{0,1}(v)= \sum_{k=1}^{N}
  \delta\tilde{\eta}_{0,1}^k\theta(v_k-v),\nonumber\\
&&\delta\tilde{\eta}_{0,1}^k\equiv
  \tilde{\eta}_{0,1}^k-\tilde{\eta}_{0,1}^{k-1},\nonumber\\
&&  \delta\tilde{\eta}_0^k>0
\label{eq:delta_eta01_piecewise}
\end{eqnarray}  

\noindent with the relations:

\begin{equation}
  \tilde{\eta}_{0,1}^k(v_{min})= \sum_{i=k}^{N}
  \delta\tilde{\eta}_{0,1}^i,
\label{eq:eta01_delta_eta01}
\end{equation}  

\noindent in a {\it large--enough} set of velocity steps (streams),
$v=[v_1,...,v_N]$. Actually, a halo function of the form
(\ref{eq:eta01_delta_eta01}) corresponds to a velocity distribution
given by $f(v)=\sum_k^N \lambda_k \delta(v-v_k)$ that has been
understood in the literature to extremize ``generalized moments'' of
the form (\ref{eq:int_r_eta}), i.e. $S=\int_{v^*}^{\infty} {\cal H}(v)
f(v,t)\,dv$ in terms of $N_c+1$ streams when $N_c$ experimental
constraints $S_k$, $k=1...N_c$ (plus the normalization of $f(v)$), are
provided\cite{gondolo_scopel}. This has been used in the literature to
minimize the likelihood function ${\cal
  L}(f,S_k)$\cite{Ibarra_Rappelt_2017,Feldstein_2014,Fox_2014} in
terms of $N \le N_c+1$ streams . Notice, however, that the likelihood
function in the analysis of Section \ref{sec:results} will depend on
the two independent halo functions $\tilde{\eta}_0$ and
$\tilde{\eta}_1$, so that in our case the number of streams N needed
to minimize $\cal L$ is not related to the number of experimental
constraints $N_c$.  However, as it is evident from
Eq.(\ref{eq:hierarchy}), the pSIDM scenario that we wish to analyze
leads to a velocity range for the DAMA modulation effect
$v_{DAMA}^{min}<v<v_{DAMA}^{max}$ which is compressed to values close
to $v_{esc}$ (typically, $v_{esc}-v_{DAMA}^{min}\le$ 50 km/sec) so in
such a small range a relatively small value of N still allows a good
sampling of the halo functions $\tilde{\eta}_{0,1}$. In fact, taking
into account the first of the requirements of
Eq.(\ref{eq:eta_conditions}), the halo function $\tilde{\eta}_0$ that
minimizes the tension between DAMA and the constraints of other
experiments (and so maximizes the likelihood function used in Section
\ref{sec:results}) is given by the minimal one that can explain the
DAMA effect, i.e. to a halo function $\tilde{\eta}_0$ monotonically
decreasing with $v_{min}$ that saturates the condition in the second
line of Eq.(\ref{eq:eta_conditions}) and flattens--out below
$v_{DAMA}^{min}$,
i.e. $\tilde{\eta}(v<v_{DAMA}^{min}$)=$\tilde{\eta}(v_{DAMA}^{min}$).
The former condition implies:

\begin{equation}
  \tilde{\eta}_{0,k}=\max_{i>k}|\tilde{\eta}_{1,i}|,
  \label{eq:eta_0_min}
\end{equation}

\noindent while the latter corresponds to:

\begin{equation}
  \delta\tilde{\eta}_{1}^k=0 \,\,\,\mbox{for $v_k<v_{DAMA}^{min}$}.
  \label{eq:minimal_eta0}
  \end{equation}

\noindent As a consequence of the considerations above, in Section
\ref{sec:results} we will adopt the $\tilde{\eta}_{1,k}$'s as free
parameters subject to (\ref{eq:minimal_eta0}) and use
(\ref{eq:eta_0_min}) for the $\tilde{\eta}_{0,k}$'s.

\noindent When the piece--wise definition of the $\eta$ function
(\ref{eq:eta01_piecewise}) is used in (\ref{eq:r_h}) one gets:

\begin{eqnarray}
R_{[E_1^{\prime},E_2^{\prime}]}&=&N_T MT
\frac{\rho_{\chi}}{m_{\chi}}\sigma \int_{v^*_{min}}^{\infty} dv
\frac{d}{dv}\left\{ v
\frac{c^2}{v}\int_{E_R^{-}(v)}^{E_R^{+}(v)}dE_R\left\{\hat{\cal
  R}_0(E_R)+\hat{\cal R}_1(E_R)[v^2-v_{min}(E_R)^2]\right\}
\right\}\nonumber\\ &&\times \sum_{k=1}^{N}
\delta\tilde{\eta}^k\theta(v_k-v)= N_T MT
\frac{\rho_{\chi}}{m_{\chi}}\sigma c^2 \sum_{k=1}^{N}
\delta\tilde{\eta}^k\times\\ &&
\int_{E_R^{-}(v)}^{E_R^{max}(v_k)}dE_R\left\{\hat{\cal
  R}_0(E_R)+\hat{\cal R}_1(E_R)[v^2_k-v_{min}(E_R)^2]\right\}.
  \end{eqnarray}

\noindent Expanding explicitly the square of
$v_{min}(E_R)$ and setting:

\begin{eqnarray}
  &&\bar{\hat{{\cal R}}}_{0,1}(E_R)\equiv \int_0^{E_R} dE^{\prime}_R \hat{{\cal R}}_{0,1}(E_R^{\prime})\nonumber\\
  &&\bar{\hat{{\cal R}}}_{1E}(E_R)\equiv \int_0^{E_R} dE^{\prime}_R E_R^{\prime} \hat{{\cal R}}_{1}(E_R^{\prime})\nonumber\\
  &&\bar{\hat{{\cal R}}}_{1E^{-1}}(E_R)\equiv \int_0^{E_R} dE^{\prime}_R \frac{1}{E_R^{\prime}} \hat{{\cal R}}_{1}(E_R^{\prime}),
  \label{eq:rbar_hat}
  \end{eqnarray}
\noindent the predicted rate can then be written as:
\begin{eqnarray}
  &&R_{[E_1^{\prime},E_2^{\prime}]}=\frac{\rho_{\chi}}{m_{\chi}}\sigma c^2 \sum_{k=1}^{N} \delta\tilde{\eta}^k\times \nonumber\\
  &&  \left \{ \bar{{\cal R}}_{0}\left [E_R^{max}(v_k) \right ]-\bar{{\cal R}}_{0}\left [E_R^{min}(v_k) \right ]+(v_k^2-\frac{\delta}{\mu_{\chi N}})
 \left ( \bar{{\cal R}}_{1}\left [E_R^{max}(v_k)\right ]-\bar{{\cal R}}_{1}\left [E_R^{min}(v_k)\right ]\right ) \right .\nonumber\\
&&   -\frac{m_T}{\mu_{\chi N}^2}
 \left ( \bar{{\cal R}}_{1E} \left [E_R^{max}(v_k)\right ]-\bar{{\cal R}}_{1E} \left [E_R^{min}(v_k)\right ]\right )\nonumber \\
 && \left . -\frac{\delta^2}{m_T}\left (\bar{{\cal R}}_{1E^{-1}} \left [E_R^{max}(v_k)\right ]-\bar{{\cal R}}_{1E^{-1}} \left [E_R^{min}(v_k)\right ]\right ) \right \},
  \label{eq:rate_rbar}
\end{eqnarray}

\noindent with:

\begin{equation}
  (\bar{{\cal R}}_{0,1},\bar{{\cal R}}_{1E},\bar{{\cal R}}_{1E^{-1}})=N_T MT (\bar{\hat{{\cal R}}}_{0,1},\bar{\hat{{\cal R}}}_{1E},\bar{\hat{{\cal R}}}_{1E^{-1}}).
  \label{eq:rbar}
  \end{equation}

\noindent The functions of Eq.(\ref{eq:rbar}) are quadratic in the two
couplings $c^n$=$c^n_4$ and $c^p$=$c^p_4$ so that, in terms of
$r\equiv\frac{c^n}{c^p}$:

\begin{equation}
  \bar{\cal R}(r)=\frac{r(r+1)}{2}\bar{\cal R}(r=1)+(1-r^2)\bar{\cal R}(r=0)+\frac{r(r-1)}{2}\bar{\cal R}(r=-1).
  \label{eq:isospin_rotation}
  \end{equation}

\begin{figure}
\begin{center}
  \includegraphics[width=0.47\columnwidth]{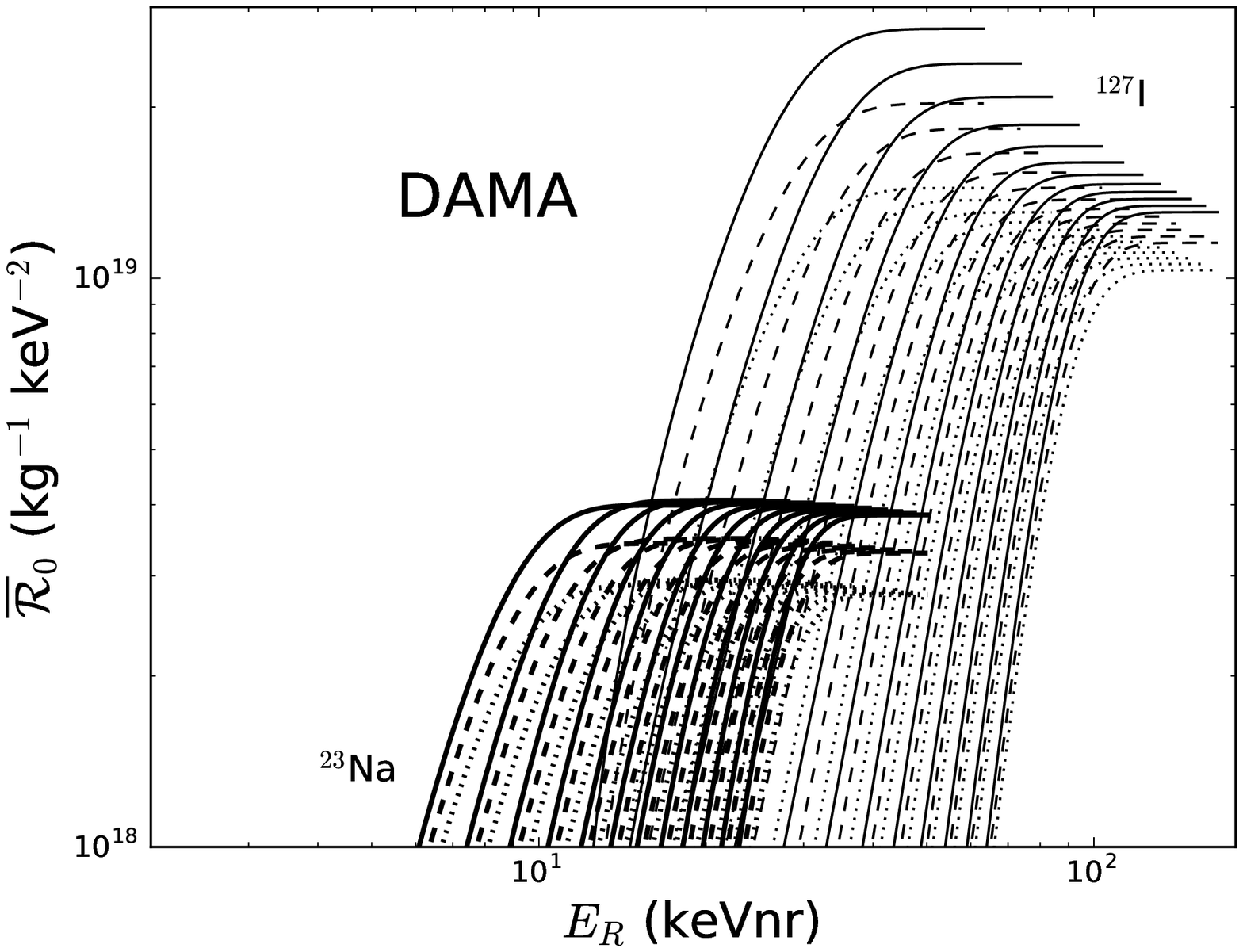}
  \includegraphics[width=0.47\columnwidth]{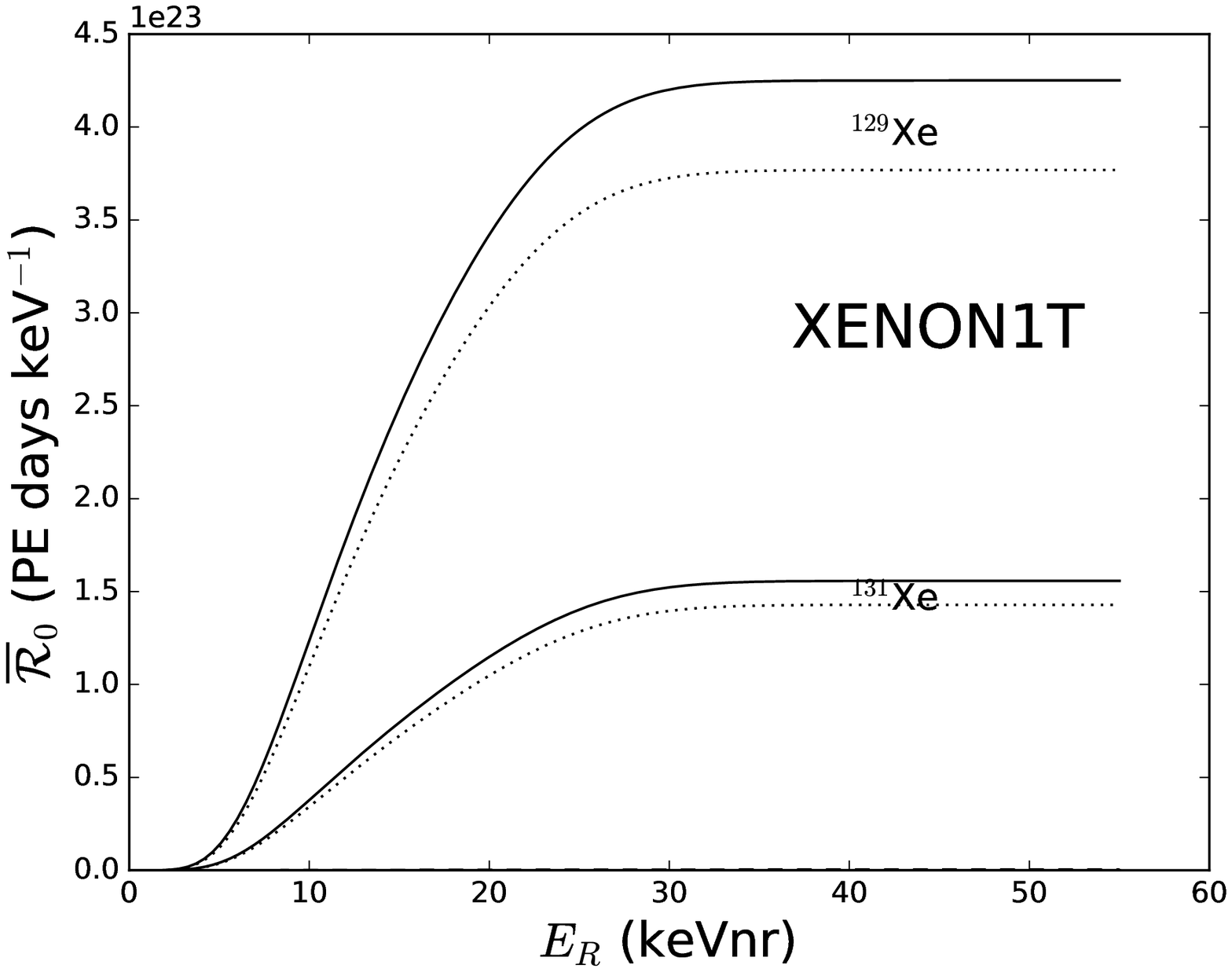}
\end{center}
\caption{Left hand plot: integrated response functions $\bar{{\cal
      R}}_0$ defined in Eqs.(\ref{eq:rbar_hat}) and (\ref{eq:rbar}) as
  a function of $E_R$ for the DAMA experiment \cite{dama_last}; right
  hand plot: the same for XENON1T\cite{xenon_1t}. As explained in
  Section \ref{sec:expected_rates} a different response function must
  be tabulated for each experimental energy bin, nuclear target
  (including different isotopes, see
  Eq.(\ref{eq:response_function_final})) and for three values of the
  coupling ratio $r=c^n/c^p$ (in order to perform the isospin rotation
  of Eq(\ref{eq:isospin_rotation})). Including $r=-1,0,1$, all
  energetic bins and target isotopes with a non--vanishing
  spin--dependent response function this implies, for instance, 72
  response functions for DAMA and 6 for XENON1T.}
\label{fig:response_functions}
\end{figure}

\noindent In the present analysis we will consider only a standard
WIMP--nucleus spin--dependent interaction with no explicit velocity
dependence in the cross section. In this case only the integrated
function $\bar{{\cal R}}_{0}$ is needed to evaluate expected rates.
For each experiment it can be tabulated as a function of $E_R$ and for
$r=-1,0,1$. Evaluations at run time for different values of
$m_{\chi}$, $\delta$ and $r$ can be obtained in a fast and efficient
way through linear combinations of $\bar{{\cal R}}_0$ interpolations
using (\ref{eq:rate_rbar}) and (\ref{eq:isospin_rotation}).  Some
examples of the functions $\bar{\cal R}_0$ are plotted in
Fig. \ref{fig:response_functions} as a function of $E_R$.

\section{Results}
\label{sec:results}

In this Section we compare the pSIDM scenario reviewed in Section
\ref{sec:spin_idm_scenario} to the current dark matter direct
detection bounds listed in Appendix \ref{app:exp} in a statistical
analysis where we wish to construct approximate 2D frequentist
confidence intervals for the set of parameters
$\theta\equiv(m_{\chi},\delta,r)$ with the velocity distribution
treated as a set of nuisance parameters $\eta$. In order to do so, in
the following we will consider both a halo--independent scenario and a
more conventional Maxwellian velocity distribution. For a given
dataset ${\bf d}$, including $N_{DAMA}$ bins for the DAMA modulation
amplitude and $i=1,...,N_{exp}$ experiments each with $N_{bin}^i$
energy bins, we construct the Likelihood function:

\begin{equation}
-2 \ln {\cal L}({\bf d}|\Theta)=\sum_{n=1}^{N_{DAMA}}\left
  (\frac{S_{1,n}(\Theta)-S_{1,n}^{exp}}{2 \sigma_n^{exp}}
\right)^2 -2\sum_{i=1}^{N_{exp}}\sum_{j=1}^{N_{bin}^i} {\cal L}_j^i(\Theta)
\label{eq:likelihood}
\end{equation}

\noindent where $\Theta=(\theta,\eta)$, with

\begin{equation}
-2 {\cal L}_j^i(\Theta)=2\left
        [S_{0,j}^i(\Theta)+B_j^i-N_j^i-N_j^i\ln\frac{S(\Theta)_{0,j}^i+B_j^i}{N_j^i}
          \right ],
\label{eq:likelihood_constraints}        
  \end{equation}

\noindent In Eq.(\ref{eq:likelihood}) $S_{1,n}$ is the prediction of
the DAMA modulation amplitude in the $n$--th bin while $S_{1,n}^{exp}$
the corresponding measurement with error $\sigma_n^{exp}$, $S_{0,j}^i$
the expected rate in the $i$--the energy bin of the $j$--th experiment
with $N_j^i$ the corresponding measured count rate and $B_j^i$ the
expected background.  As far as the background is concerned, we notice
that with the current level of required sensitivities its estimation
is subject to large uncertainties. For this reason we assume the
$B_j^i$'s as free parameters and minimize the likelihood with respect
to them.  This corresponds to taking:

\begin{equation}
-2 {\cal L}_j^i(\Theta)=\left\{
  \begin{array}{ll}
2\left
        [S_{0,j}^i(\Theta)-N_j^i-N_j^i\ln\frac{S(\Theta)_{0,j}^i}{N_j^i}
          \right ] & \mbox{if $S_{0,j}^i(\Theta)>N_j$ } \\
        0 & \mbox{otherwise}.
    \end{array}
  \right . 
  \end{equation}

\noindent In the case of the PICASSO experiment\cite{picasso} for each
energy threshold $E^{th}_i$ the number of observed events $x_i^{exp}$
and 1--sigma Gaussian fluctuation $\sigma_i$ normalized to
events/kg/day is provided (see Table \ref{table:picasso} in Appendix
\ref{app:exp}). To include such constraint in our analysis we modify
the likelihood function in the following way:

\begin{equation}
-2 \ln {\cal L}({\bf d}|\Theta) \rightarrow -2 \ln {\cal L}({\bf
  d}|\Theta)+\sum_i\left (\frac{x_i-x_i^{exp}}{\sigma_i} \right
)^2\Theta(x_i-x_i^{exp})
\label{eq:likelihood_picasso}
  \end{equation}

\noindent with $x_i$ the theoretical prediction of the corresponding
detection rate.

We then construct approximate 2D and 1D frequentist confidence
intervals for model parameters from an effective chi-square defined as
$\Delta \chi^2_{eff}\equiv -2 \ln {\cal L}_{prof}/{\cal L}_{max}$
where ${\cal L}_{max}$ is the maximum likelihood and:

\begin{equation}
  {\cal L}_{prof}({\bf d},\theta_1)\propto
  \max_{\theta_2,\theta_3,\eta} {\cal L}({\bf
    d}|\Theta)\,\,\, \mbox{or}\,\,\,  {\cal L}_{prof}({\bf d},\theta_1,\theta_2)\propto
  \max_{\theta_3,\eta} {\cal L}({\bf
    d}|\Theta),
  \label{eq:eff_chi2}
  \end{equation}

\noindent Wilks' theorem guarantees that under certain regularity
conditions the corresponding distributions of $\Delta \chi^2_{eff}$
converge to a chi-square distribution with 1 or 2 degrees of freedom
\cite{profile_likelihood}.

The practical implementation of the method described above is
conceptually quite simple but may require to explore a parameter space
of large dimensionality (at least in the halo--independent approach).
This kind of task is efficiently performed by using the technique of
Markov chains, which makes use of the likelihood function itself to
optimize the sampling procedure. To this aim we use the Markov--Chain
Monte Carlo (MCMC) code emcee~\cite{emcee} to generate a large number
of sets $(\theta,\eta)$.

\subsection{The halo--independent case}
\label{sec:halo-independent}

\begin{figure}
\begin{center}
  \includegraphics[width=0.32\columnwidth]{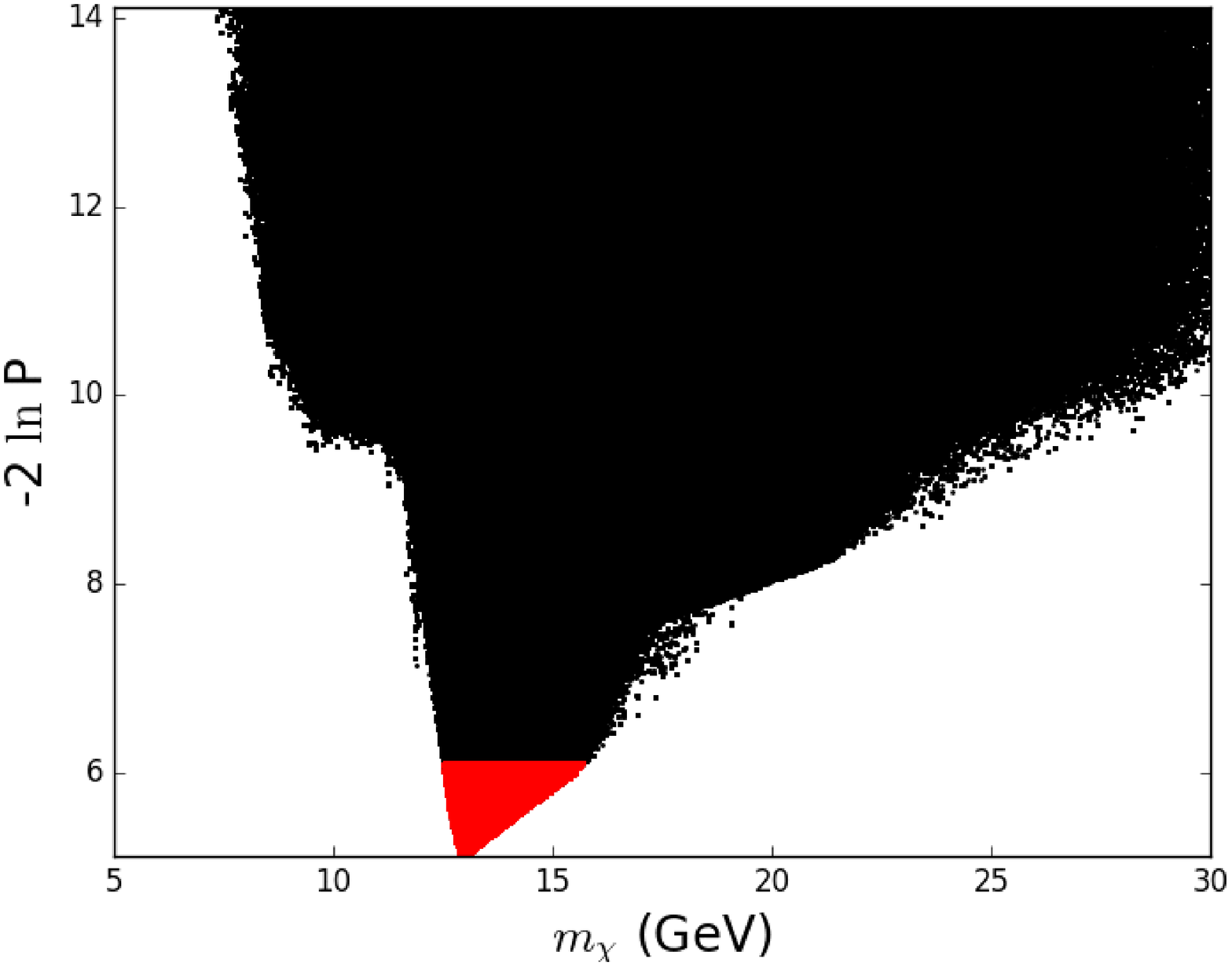}
  \includegraphics[width=0.32\columnwidth]{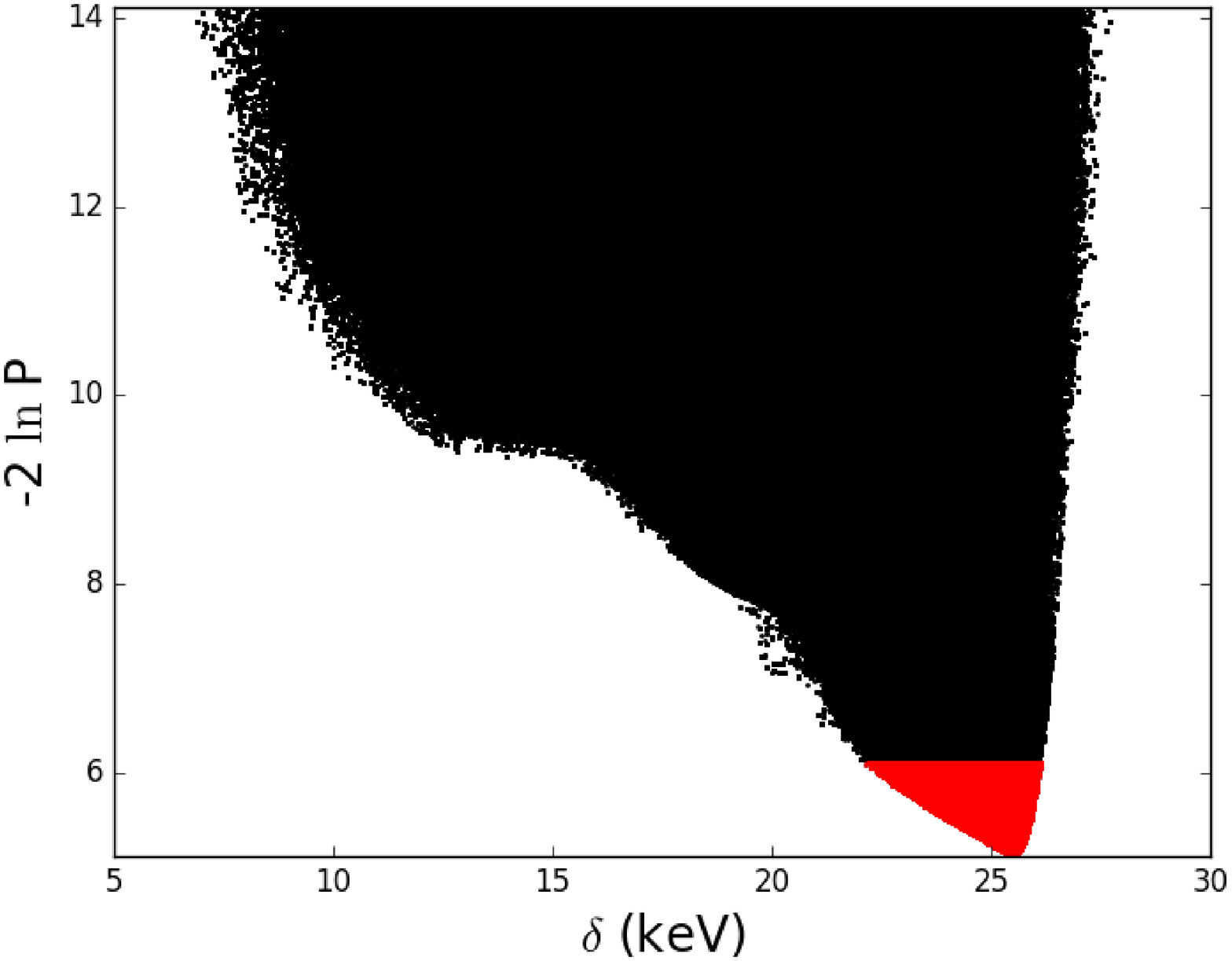}
  \includegraphics[width=0.32\columnwidth]{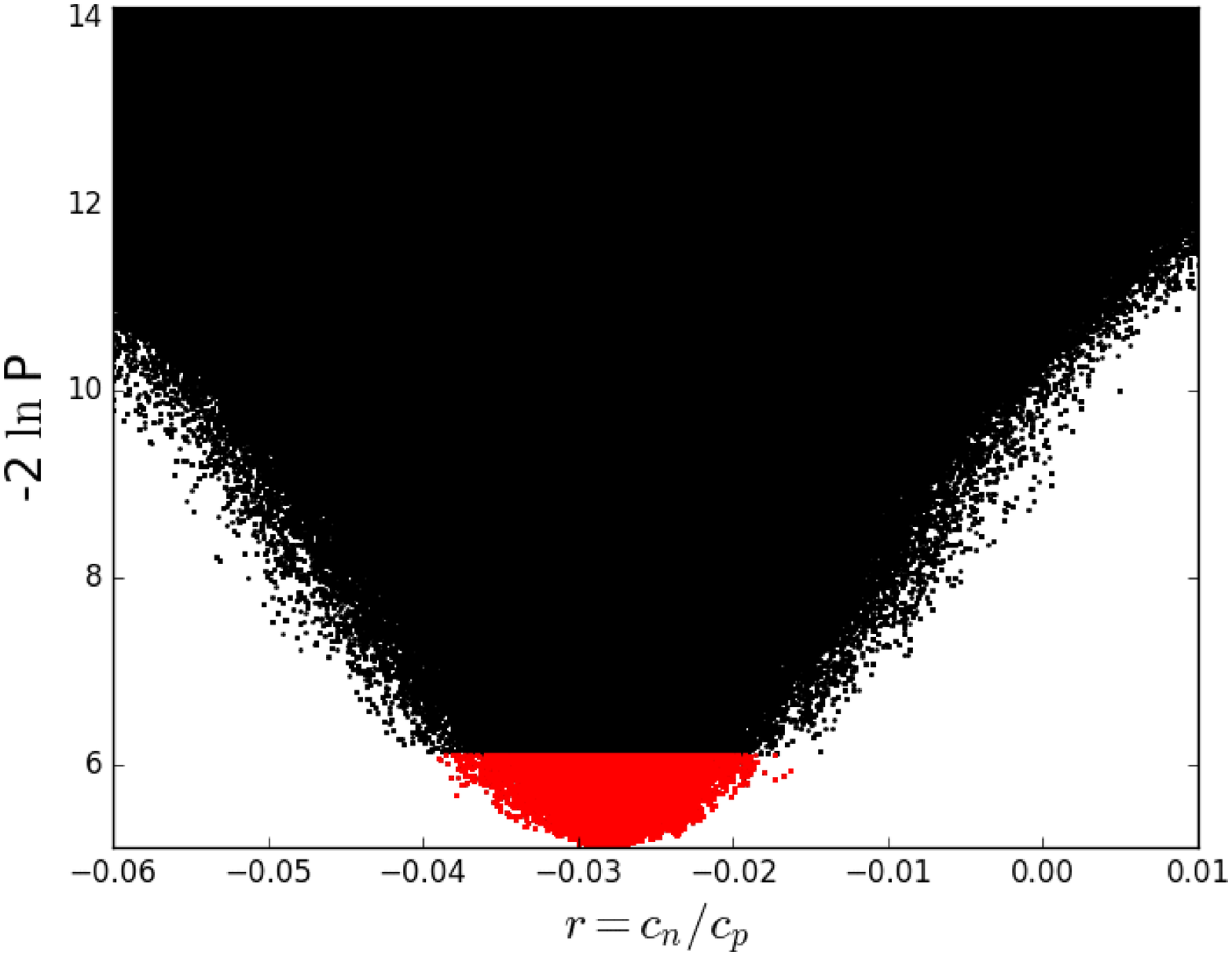}  
\end{center}
\caption{Profile likelihood of the WIMP mass $m_{\chi}$ (left) the
  mass splitting $\delta$ (center) and of the coupling ratio $r$
  (right) when the generalized halo functions $\tilde{\eta}_{0,1}$ are
  parameterized in terms of Eqs.(\ref{eq:delta_eta01_piecewise},
  \ref{eq:minimal_eta0}, \ref{eq:eta_0_min}). Red points are subject
  to the condition (\ref{eq:1sigma_condition}.)}
\label{fig:profiles}
\end{figure}

\begin{figure}
\begin{center}
  \includegraphics[width=0.32\columnwidth]{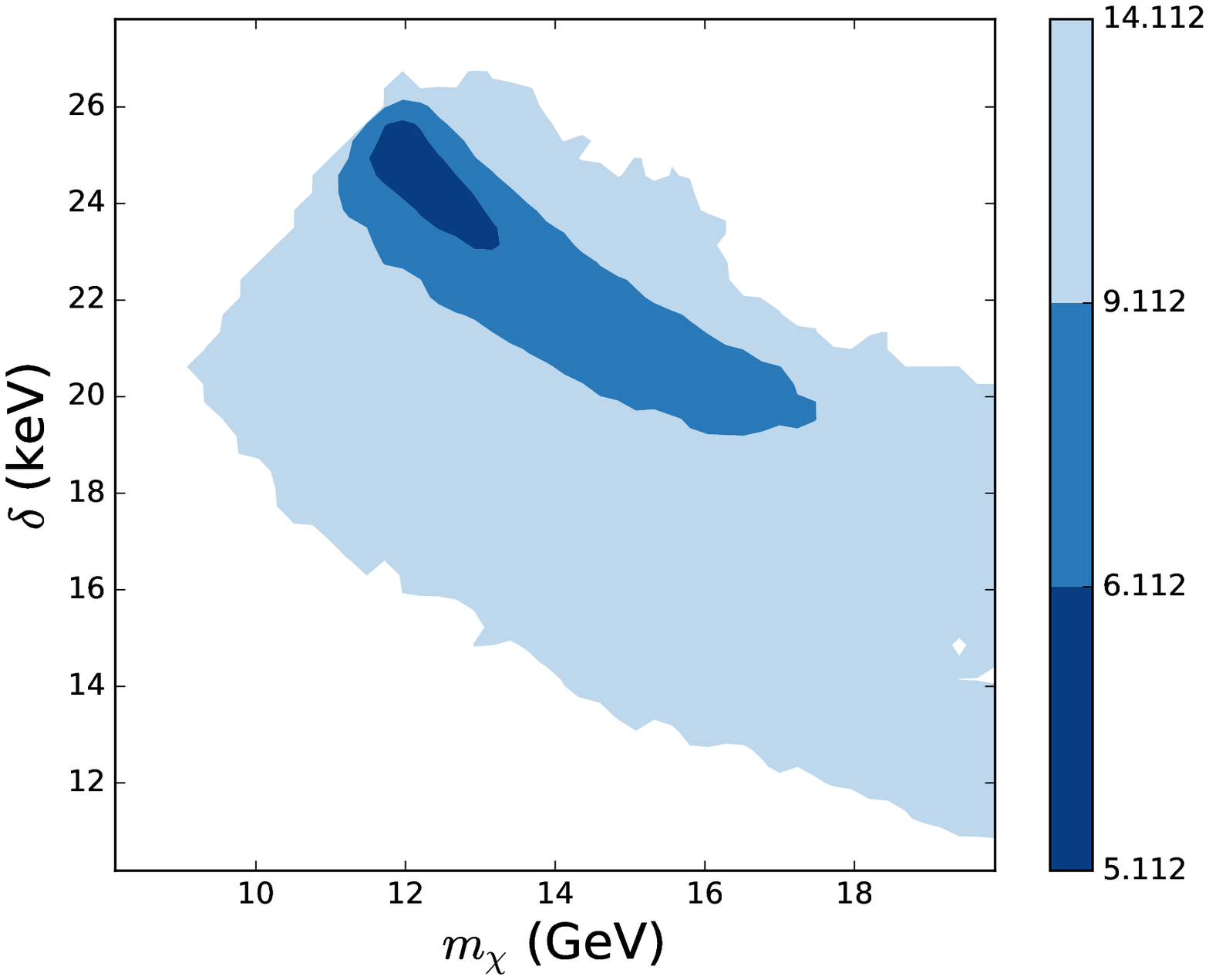}
  \includegraphics[width=0.32\columnwidth]{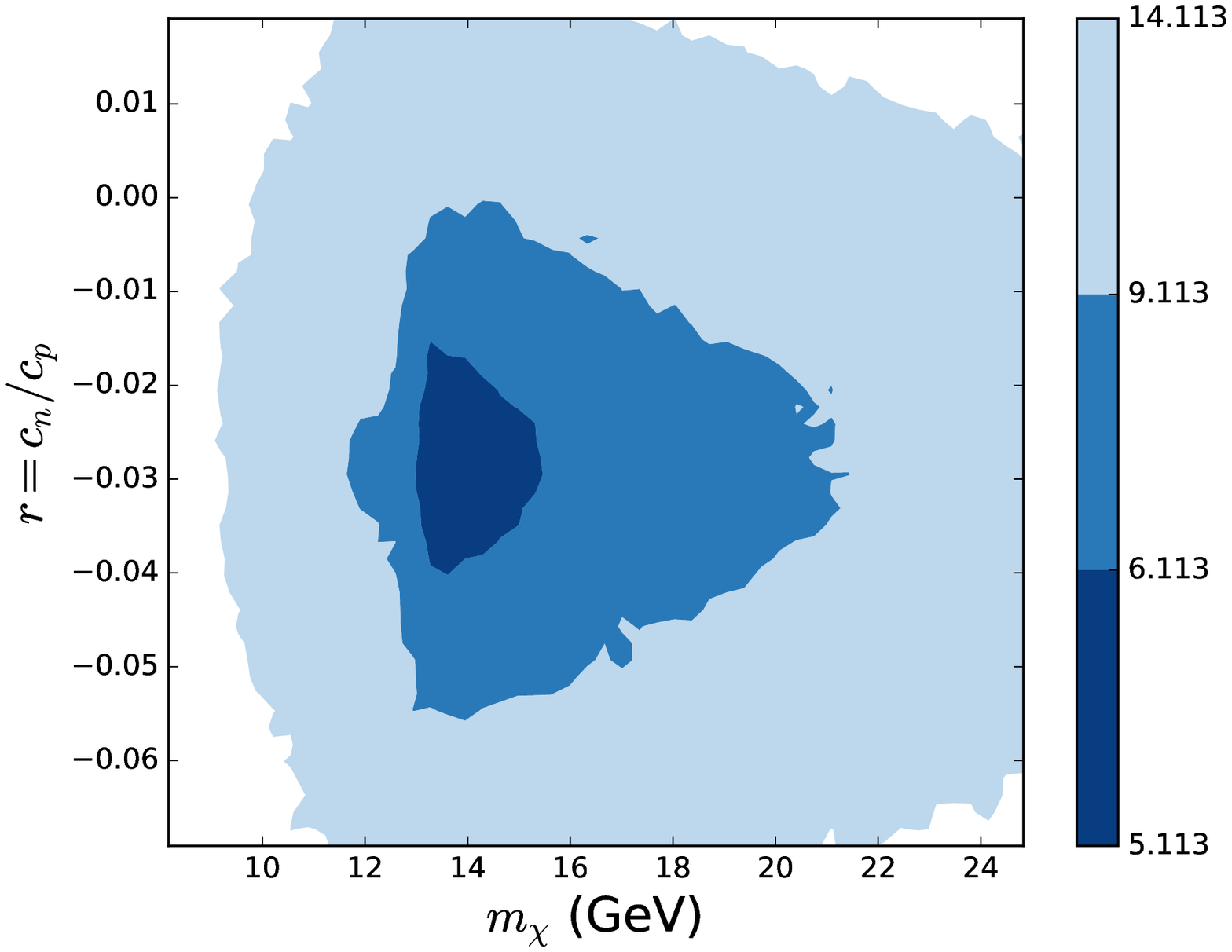}
  \includegraphics[width=0.32\columnwidth]{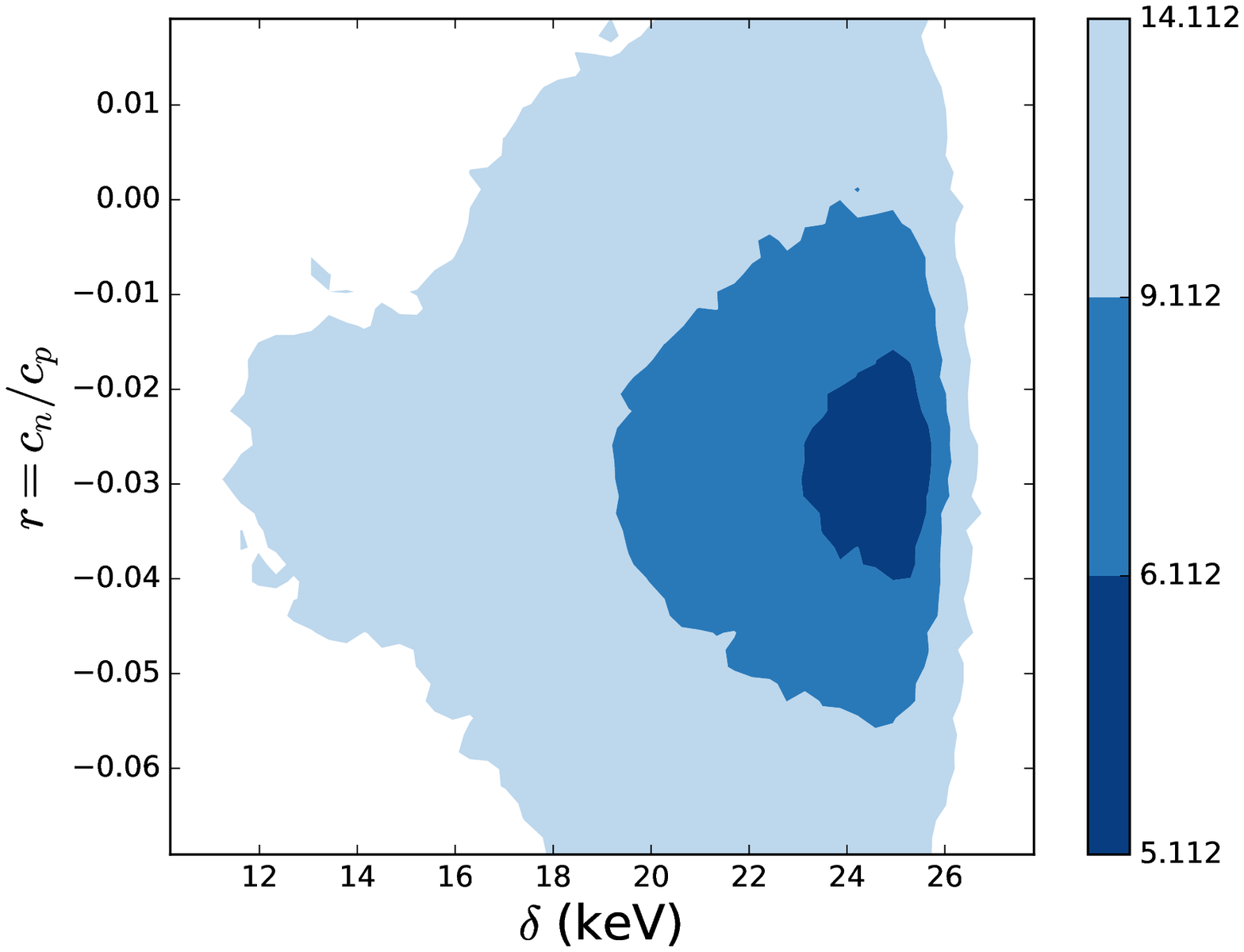}  
\end{center}
\caption{Correlations among the WIMP mass $m_{\chi}$, the mass
  splitting $\delta$ and of the coupling ratio $r$ plotted in
  Fig.(\ref{fig:profiles}). Dark shade: $-2\ln P<(-2\ln P)_{min}+1$;
  medium shade: $-2\ln P<(-2\ln P)_{min}+4$; light shade: $-2\ln
  P<(-2\ln P)_{min}+9$.}
\label{fig:correlations}
\end{figure}

In the halo--independent approach we parameterize the halo function
$\tilde{\eta}_{1}$ with Eq.(\ref{eq:delta_eta01_piecewise}) subject to
the conditions (\ref{eq:eta_conditions}) and (\ref{eq:minimal_eta0}),
and express $\tilde{\eta}_{0}$ using (\ref{eq:eta_0_min}), i.e. in
terms of the {\it minimal} halo function that can explain the DAMA
effect. In this way
$\Theta=(\theta,\eta)=(m_{\chi},\delta,r,v_i,\delta\tilde{\eta}_1^i)$
with $\theta=(m_{\chi},\delta,r)$, $\eta=(v_i,\delta\tilde{\eta}_1^i)$
and $i=1,...,N$. As already pointed out, while in general the number
of streams $N$ would need to be arbitrarily large, in the pSIDM
scenario moderate $N$ values are expected to sample the halo functions
well enough, since one has $v_{min}^*<v_i<v_{esc}$ with
$v_{min}^*\rightarrow v_{esc}$, so that the range of $v_i$ is
particularly compressed (for $v_{esc}$ in this Section we assume
$v_{esc}$=782 km/s, which is obtained by combining a WIMP escape
velocity $u_{esc}$ =550 km/s in the Galactic rest frame
\cite{vesc_2014} with a rotational velocity of the Solar system
$v_0$=220 km/s \cite{v0_koposov} and a small peculiar velocity
component, i.e.  $v_{esc}=u_{esc}+v_0+12$). Indeed, we have found that
numerically the absolute maximum of the likelihood is achieved for
$N$=4 (-2 $\ln {\cal L}\simeq 5.1)$ and that the profile likelihood
does not improve for larger values of $N$. In our results we combine
$N=2,..,8$ and for each value of $N$ we generate a Markov chain of
$5\times10^6$ points using 250 independent walkers and a standard
Metropolis-Hastings sampler, for a total of $3.5\times 10^7$
points. The outcome of this analysis is provided in
Fig. \ref{fig:profiles}, where the effective chi-square of
Eq.(\ref{eq:likelihood}) is plotted versus the three pSIDM parameters
$\theta_i=(m_{\chi},\delta,r)$. Their 1$\sigma$ confidence intervals
can then be obtained through the condition:

\begin{equation}
\Delta(-2 \ln {\cal L}_{prof}(\theta_i))\equiv -2 \ln {\cal
  L}_{prof}(\theta_i)+ 2 \ln {\cal L}_{max} \le 1,
\label{eq:1sigma_condition}
\end{equation}

\noindent with ${\cal L}_{prof}(\theta_i)$ given by
Eq.(\ref{eq:eff_chi2}). In Fig. \ref{fig:profiles} the corresponding
points are plotted in red. Notice that these 1$\sigma$ confidence
intervals, sometime called 1$\sigma$ likelihood intervals, have a 68\%
coverage probability in the limit of large samples when the likelihood
is well approximated by a Gaussian, but do not necessarily have a
coverage probability of 68\% if the likelihood is non-Gaussian, as is
likely the case from inspection of
Fig. \ref{fig:profiles}. Correlations between couples of the
parameters $(m_{\chi},\delta,r)$ can be obtained by plotting contour
plots of $\Delta(-2 \ln {\cal L}_{prof}(\theta_i))$: the thee
corresponding plots are shown in Fig.(\ref{fig:correlations}), with
1--$\sigma$ likelihood intervals in dark blue.

Quantitatively, in the halo--independent case the 1--$\sigma$
likelyhood intervals for the pSIDM parameters turn out to be:

\begin{eqnarray}
  &&      \mbox{12.5 GeV} \le m_{\chi}\le \mbox{15.7 GeV}  \nonumber \\
  &&       \mbox{22.1 keV}  \le \delta \le \mbox{26.1 keV} \nonumber \\
  &&       -0.039   \le r \le -0.016.
  \label{eq:1sigma_int_halo_indep}
\end{eqnarray}

\begin{figure}
\begin{center}
  \includegraphics[width=0.49\columnwidth]{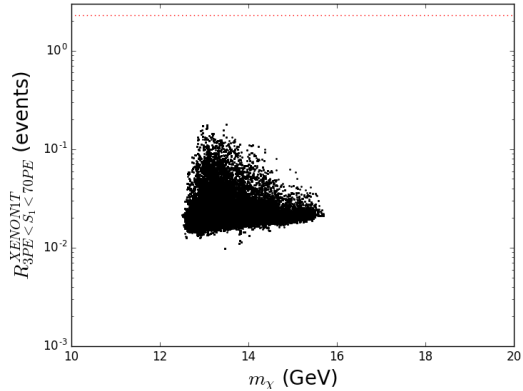}
\end{center}
\caption{The expected rate in the upper half of the
  nuclear recoil band of the $S_1$ (primary scintillation) full region
  of interest of XENON1T, corresponding to 2 PE$<S_1<$70 PE (see
  Fig. 2 of Ref.\cite{xenon_1t}) is plotted vs. the WIMP mass
  $m_{\chi}$ for the pSIDM configurations corresponding to the
  1--$\sigma$ intervals of Eq.(\ref{eq:1sigma_int_halo_indep}) and
  Figs. \ref{fig:profiles} and \ref{fig:correlations}. The horizontal
  line represents the 90\% C.L. upper bound for no nuclear--recoil
  candidates. }
\label{fig:rate_xenon_1t}
\end{figure}

\noindent The fact that the value $r$=0 is $\simeq$ 2.4 $\sigma$ away
from the best-fit value signals tension with constraints from the
neutron--odd targets $^{129}Xe$ and $^{131}Xe$ in XENON1T and PANDA.
Indeed, the value $r=c^n/c^p\simeq$=-0.03 corresponds to a
cancellation in the xenon spin--dependent nuclear form factors for
which, as already pointed out, we are using the determination from
\cite{haxton1,haxton2}\footnote{Two other determinations of the same
  form factors exist in the literature, Bonn--A and
  Nijmegen\cite{spin_form_factors}, for which the cancellation is at
  even lower values of $r$: $r\simeq$=-0.05 for the former and
  $r\simeq$=-0.08 for the latter.}. In light of this in
Fig. \ref{fig:rate_xenon_1t} we show the expected rate in XENON1T in
the upper half of the nuclear recoil band of the $S_1$ (primary
scintillation) full region of interest (corresponding to 2 PE$<S_1<$70
PE, see Fig. 2 of Ref.\cite{xenon_1t}) as a function of the WIMP mass
$m_{\chi}$ for the pSIDM configurations corresponding to the
1--$\sigma$ intervals of Eq.(\ref{eq:1sigma_int_halo_indep}) and
Figs. \ref{fig:profiles} and \ref{fig:correlations}. In the same plot
the horizontal dashed line at 2.3 represents the 90\% C.L. upper bound
on the count rate corresponding to zero observed nuclear recoil
candidates. From this plot one can see that with the current
exposition (1.78$\times 10^4$ kg day) the expected rate upper bound is
approximately 0.2, i.e. one order of magnitude below the horizontal
line. Since the first run of XENON1T was limited to only 34.2 live
days\cite{xenon_1t}, we conclude that one full year of data taking
should allow XENON1T to reach the level of sensitivity required to
start probing the pSIDM scenario.

\subsection{The Maxwellian case}
\label{sec:maxwellian}

The most natural assumption for the thermalized component of the WIMP
velocity distribution is a Maxwellian at rest in the Galactic rest
frame with a cut--off corresponding to the escape velocity. In this
case the halo functions $\eta_{0,1}(v)$ are known, while to calculate
the $\tilde{\eta}_{0,1}(v)$'s the cross--section $\sigma_p$ is an
additional free parameter. Expected rates can be calculated by
directly performing the velocity integrals (\ref{eq:int_r_eta}), an
operation that is however time consuming when performed in a MCMC
sampling. For this reason we keep using Eq.(\ref{eq:rate_rbar}) to
calculate expected rates, with the coefficients
$\tilde{\eta}_{0,1}^k$, corresponding to averages of the halo
functions $<\eta_{0,1}>_{[v_{k-1},v_k]}$ in the velocity intervals
$v_{k-1}<v<v_k$, calculated analytically. Specifically, indicating in
the Galactic rest frame with $v_{rms}$ the r.m.s. velocity of the
Maxwellian, $v_{Earth}$=$v_{Sun}$+$\Delta v_{Earth}
cos[\frac{2\pi}{T}(t-t_0)]$ the velocity of the Earth and with
$u_{esc}$ the escape velocity, they are given by:

\begin{equation}
\tilde{\eta}_{0,1}^k=\frac{\bar{\eta}_{0,1}(v_k)-\bar{\eta}_{0,1}(v_{k-1})}{v_k-v_{k-1}}
  \end{equation}

\noindent with:

\begin{equation}
  \bar{\eta}_0(v)=\frac{N}{\eta \sqrt{\pi}}\left\{
  \begin{array}{ll}
    \frac{\sqrt{\pi}}{2}\left[(x+\eta)\erf(x+\eta)-(x-\eta)\erf(x-\eta) \right ]+&\\
    +\frac{1}{2}\left [e^{-(x+\eta)^2}-e^{-(x-\eta)^2} \right ]-2 x \eta e^{-z^2},     & \mbox{if $x<z-\eta$}\\
    \frac{\sqrt{\pi}}{2}\left[(x+\eta)\erf(z)-(x-\eta)\erf(x-\eta) \right ]+     &\\
    \frac{e^{-z^2}}{2}\left (\eta^2-2\eta x-2\eta z+x^2-2 x z+1+z^2 \right )-
    \frac{1}{2}e^{(x-\eta)^2},                       &\mbox{if $z-\eta<x<z+\eta$}\\
    \sqrt{\pi}\eta \erf(z)-2\eta z e^{-z^2},       &  \mbox{$x>z+\eta$}
    \end{array}
  \right . 
  \end{equation}

\noindent and:

\begin{equation}
  \bar{\eta}_1(v)=\frac{\Delta\eta}{\eta}\left(\delta\bar{\eta}_0(v)
  -\bar{\eta}_0(v) \right)
  \end{equation}
\noindent with:

\begin{equation}
  \delta\bar{\eta}_0(v)=\frac{N}{\eta \sqrt{\pi}}\left\{
  \begin{array}{ll}
    \frac{\sqrt{\pi}}{2}\left[\erf(x+\eta)+\erf(x-\eta) \right ]
    -2 x \eta e^{-z^2},     & \mbox{if $x<z-\eta$}\\
    \frac{\sqrt{\pi}}{2}\left[\erf(z)+\erf(x-\eta) \right ]
    +(\eta-x-z)e^{-z^2},                       &\mbox{if $z-\eta<x<z+\eta$}\\
    \sqrt{\pi} \erf(z)-2 z e^{-z^2},       &  \mbox{$x>z+\eta$}
    \end{array}
  \right . 
  \end{equation}

\noindent where $x=\sqrt{3/2} v/v_{rms}$, $\eta=\sqrt{3/2}
v_{Sun}/v_{rms}$, $\Delta\eta=\sqrt{3/2}\Delta v_{Earth}/v_{rms}$
$z=\sqrt{3/2} u_{esc}/v_{rms}$ and $N=[\erf(z)-2/\sqrt{\pi}z
  e^{-z^2}]^{-1}$.  In the isothermal sphere model hydrothermal
equilibrium between the WIMP gas pressure and gravity is assumed,
leading to $v_{rms}$=$\sqrt{3/2}v_0$ with $v_0$ the galactic
rotational velocity, while $v_{Sun}$=$v_0$+12, accounting for a
peculiar component. To evaluate
Eq.(\ref{eq:rate_rbar}) in our MCMC we have adopted $N$=50.

\begin{figure}
\begin{center}
  \includegraphics[width=0.32\columnwidth]{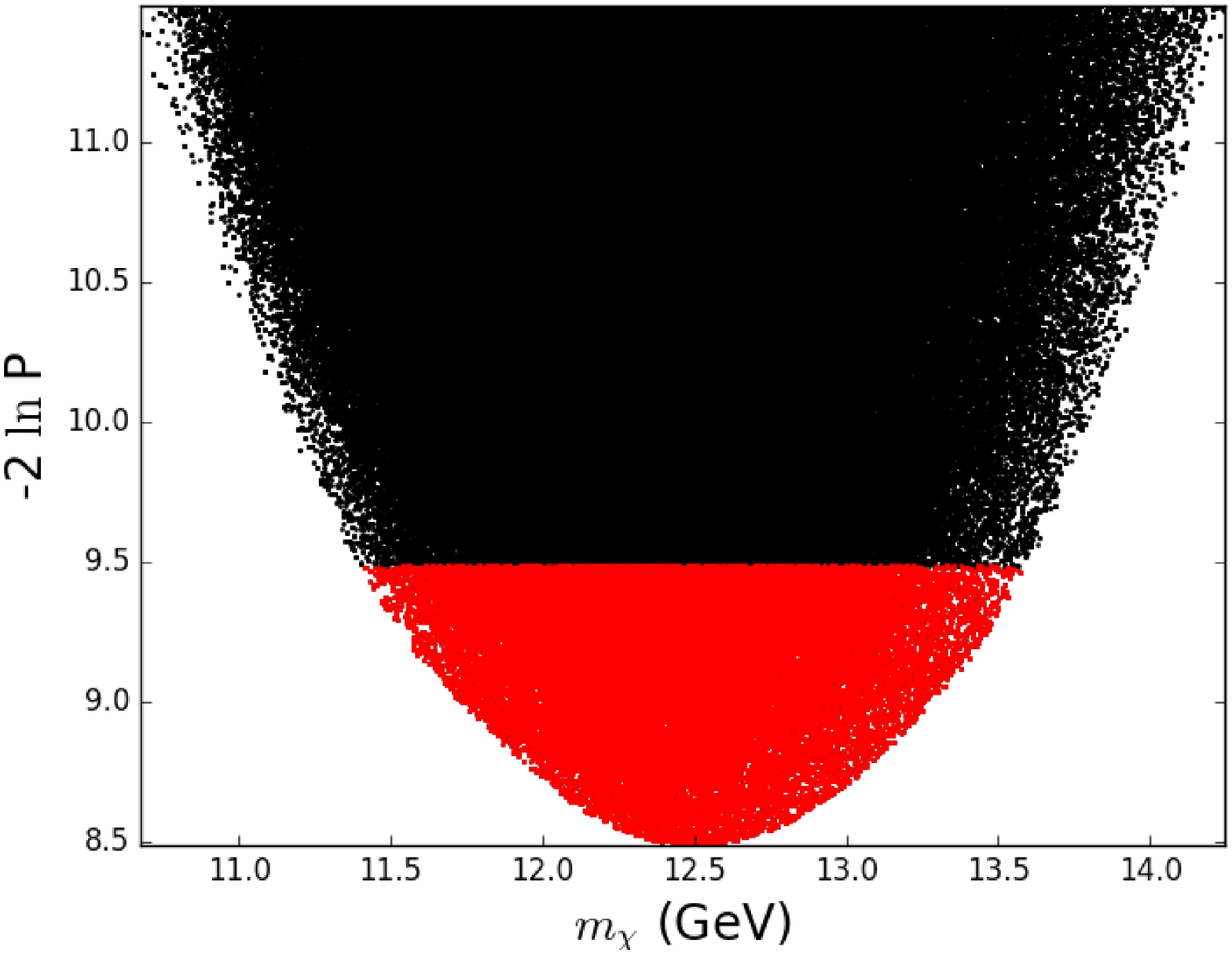}
  \includegraphics[width=0.32\columnwidth]{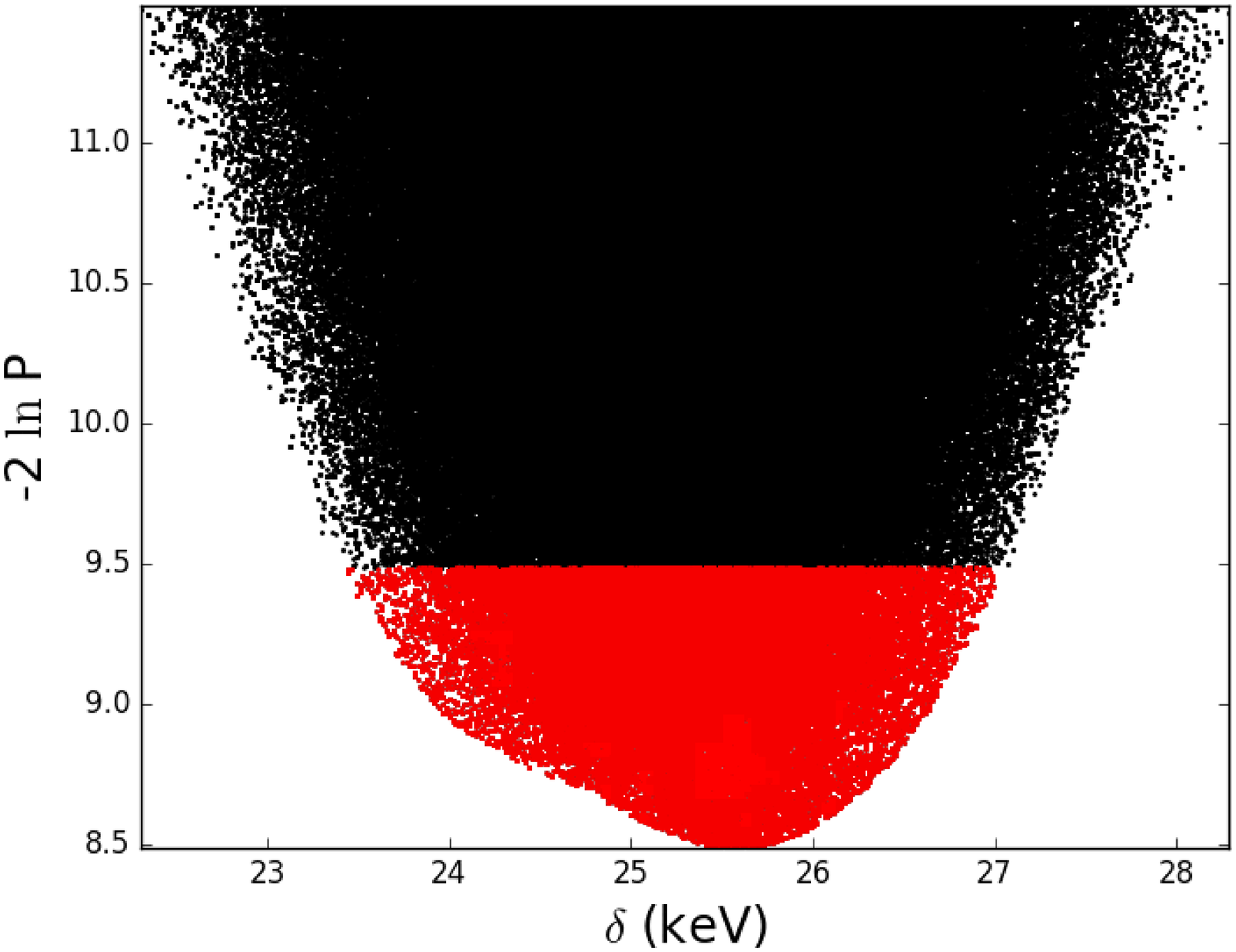}
  \includegraphics[width=0.32\columnwidth]{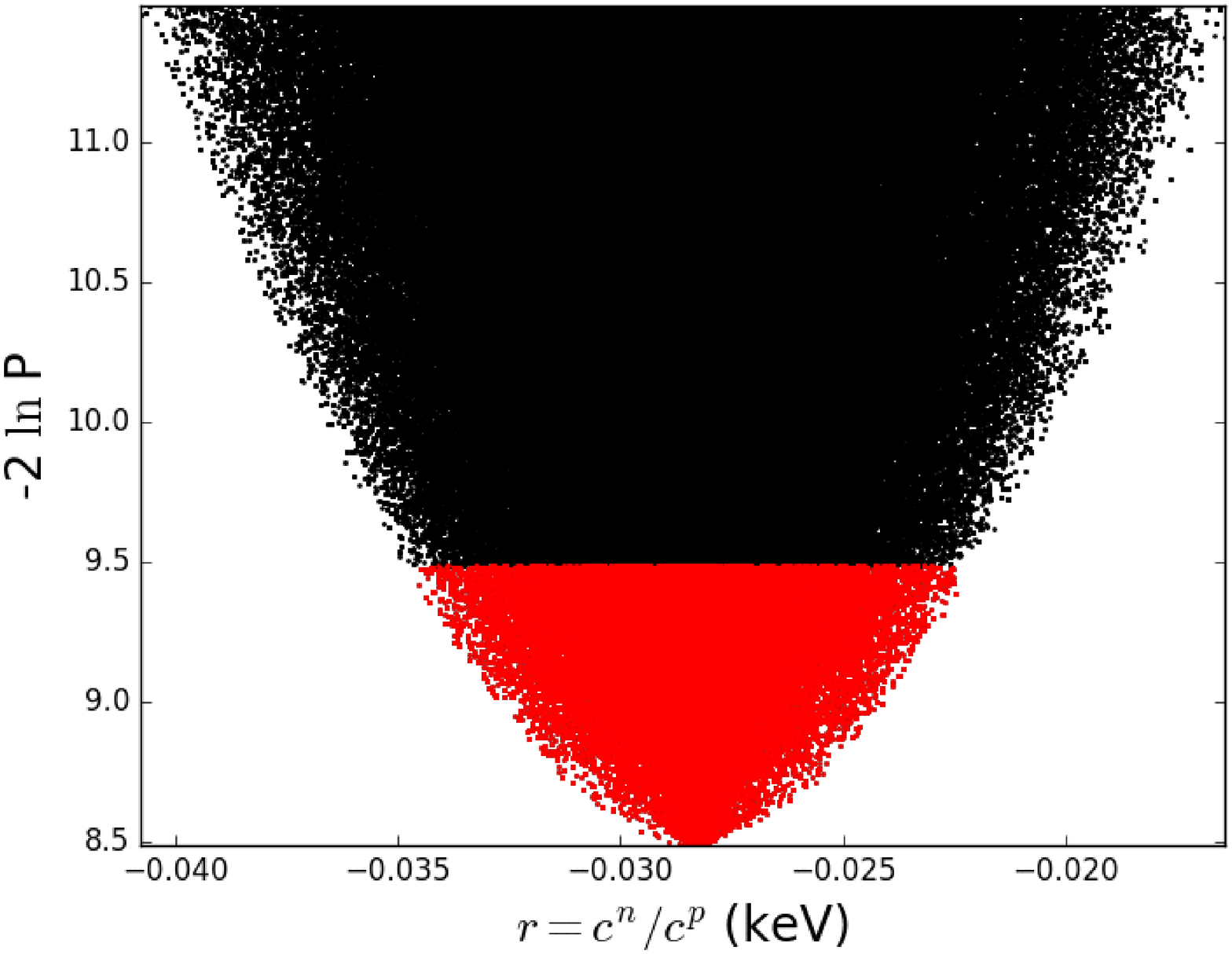}  
\end{center}
\caption{Profile likelihood of the WIMP mass $m_{\chi}$ (left) the
  mass splitting $\delta$ (center) and of the coupling ratio $r$
  (right) for a Maxwellian distribution with $v_0$=(220$\pm$ 20) km/s
  and $u_{esc}$=(550$\pm$ 30) km/s (assuming for both a corresponding
  Gaussian fluctuation in the likelihood). The color code is the same
  as in Fig.\ref{fig:profiles}.}
\label{fig:profiles_maxwellian}
\end{figure}

\begin{figure}
\begin{center}
  \includegraphics[width=0.32\columnwidth]{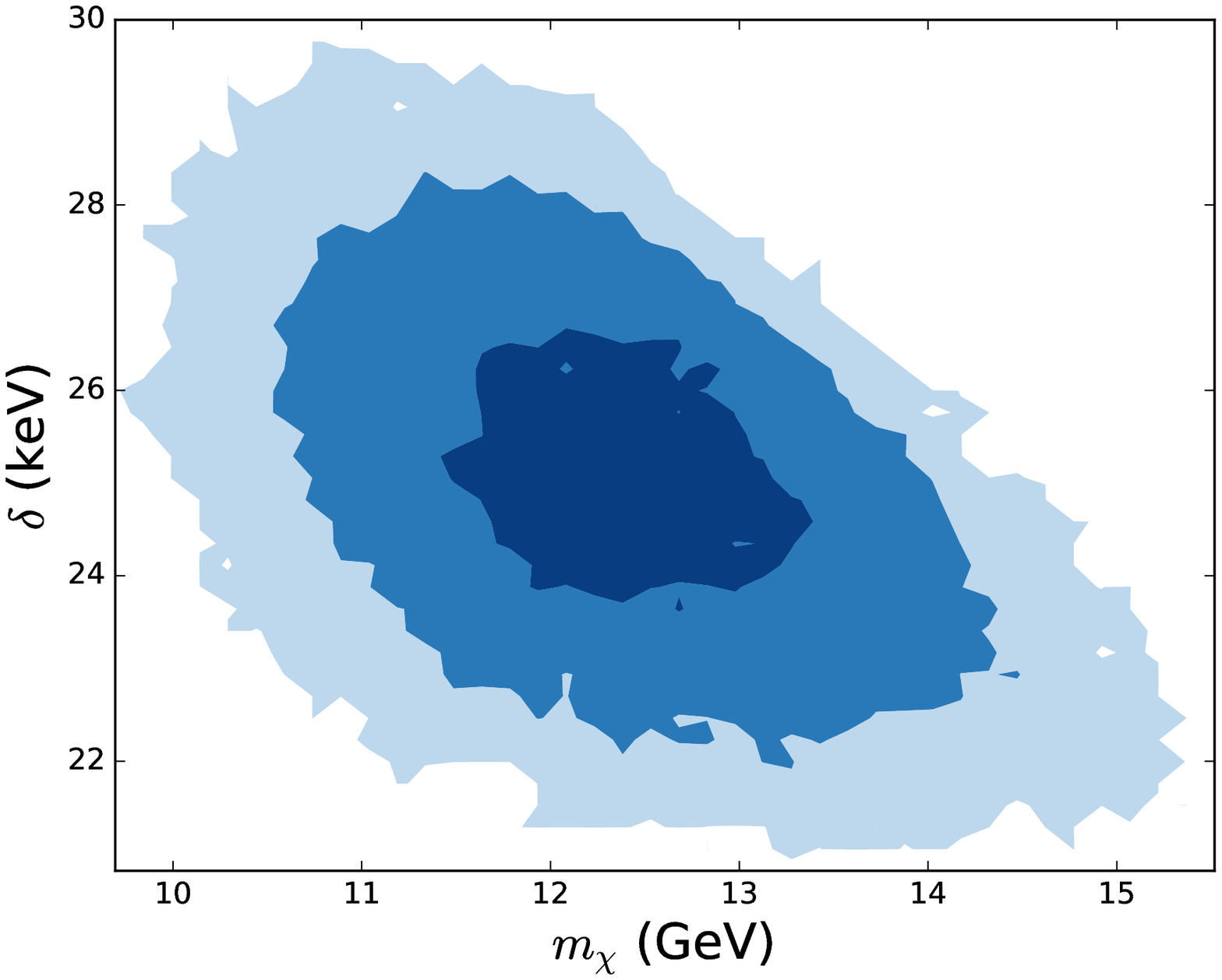}
  \includegraphics[width=0.32\columnwidth]{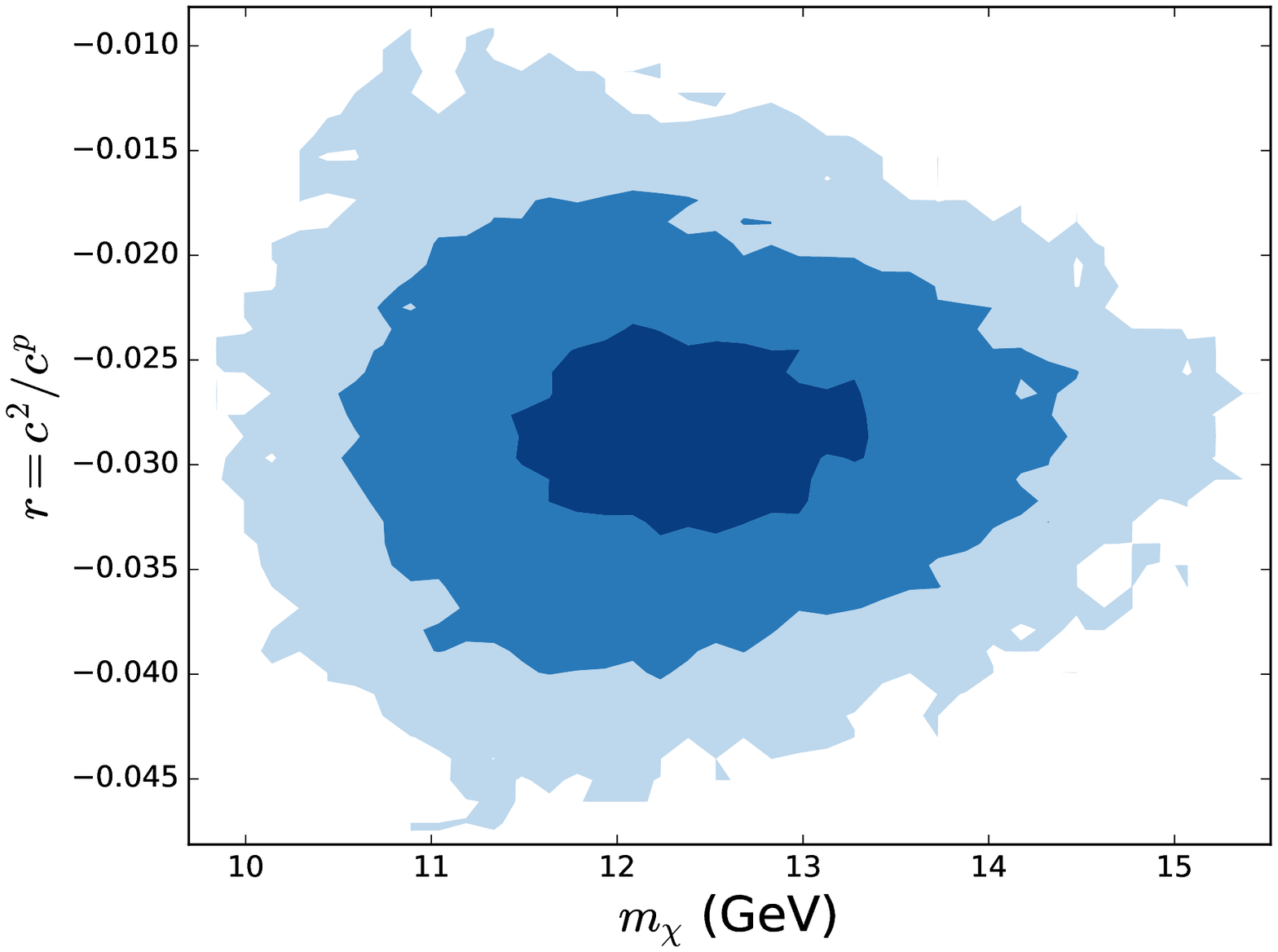}
  \includegraphics[width=0.32\columnwidth]{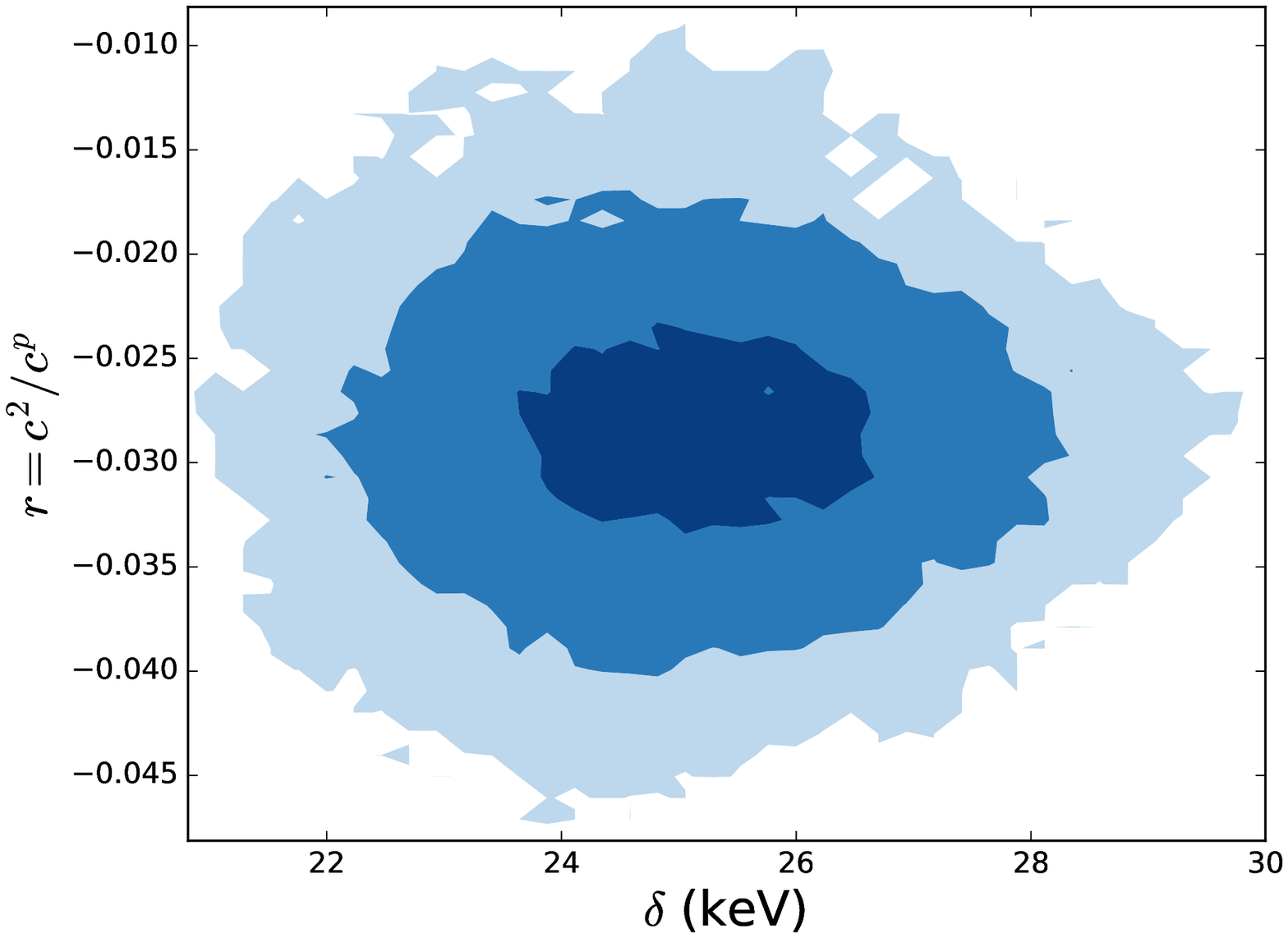}  
\end{center}
\caption{Correlations among the WIMP mass $m_{\chi}$, the mass
  splitting $\delta$ and of the coupling ratio $r$ for a Maxwellian
  distribution with $v_0$=(220$\pm$ 20) km/s and $u_{esc}$=(550$\pm$
  30) km/s (assuming for both a corresponding Gaussian fluctuation in
  the likelihood). The color codes is the same as in
  Fig.\ref{fig:correlations}.}
\label{fig:correlations_maxwellian}
\end{figure}

\begin{figure}
\begin{center}
  \includegraphics[width=0.8\columnwidth]{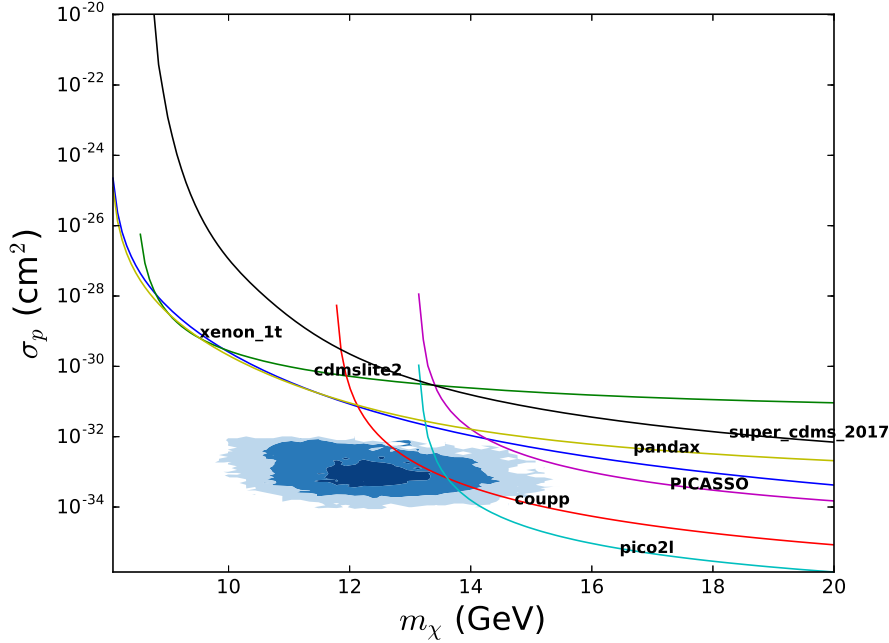}
\end{center}
\caption{Two--dimensional profile of the WIMP mass $m_{\chi}$ vs. the
  WIMP--proton conventional cross section $\sigma_p$ defined in
  Eq.(\ref{eq:conventional_sigma}) for a Maxwellian distribution with
  $v_0$=(220$\pm$ 20) km/s and $u_{esc}$=(550$\pm$ 30) km/s (assuming
  for both a corresponding Gaussian fluctuation in the
  likelihood). The color code is the same as in
  Fig.\ref{fig:profiles}. Superimposed, a standard set of
  90\% C.L. exclusion plots where each experiment is analyzed
  independently. For all exclusion plots $\delta$=25.6 keV and
  $r$=-0.028, which correspond to the absolute maximum of the
  likelihood function shown in Fig. \ref{fig:profiles_maxwellian}.}
\label{fig:mchi_sigma_maxwellian}
\end{figure}

In the Maxwellian case an additional free parameter of the model is
the WIMP--proton reference cross section $\sigma_p$ introduced in
Eq.(\ref{eq:conventional_sigma}), which represents a normalization
factor for the experimental response functions. In particular we
factorize $\sigma_p$ by fixing the WIMP local density to the standard
value $\rho_{\chi}$=0.3 GeV/cm$^3$.  For the two parameters $v_0$ and
$u_{esc}$ we take $v_0$=(220$\pm$ 20) km/s \cite{v0_koposov} and
$u_{esc}$=(550$\pm$ 30) km/s \cite{vesc_2014} assuming for both a
Gaussian fluctuation in the likelihood, i.e. adding to latter the
terms $[(v_0-220)/20]^2+[(u_{esc}-550)/30]^2$. As a consequence, we
adopt the 6 parameters $\Theta=(\theta,\eta)$,
$\theta=(\sigma_p,m_{\chi},\delta,r)$, $\eta=(v_0,u_{esc})$. The
results which correspond to such analysis are provided in
Figs. \ref{fig:profiles_maxwellian}, \ref{fig:correlations_maxwellian}
and \ref{fig:mchi_sigma_maxwellian}.  Quantitatively, in this case the
1--$\sigma$ likelyhood intervals for the pSIDM parameters turn out to
be:


\begin{eqnarray}
  &&      \mbox{3.54$\times 10^{-34}$ cm$^2$} \le \sigma_p\le   \mbox{4.09$\times 10^{-33}$ cm$^2$} \nonumber \\  
  &&      \mbox{11.4 GeV} \le m_{\chi}\le  \mbox{13.6 GeV} \nonumber \\
  &&        \mbox{24.4 keV}  \le \delta \le  \mbox{27.0 keV} \nonumber \\
  &&          -0.035    \le r \le -0.022.
  \label{eq:1sigma_int_maxwellian}
\end{eqnarray}


\section{Conclusions}
\label{sec:conclusions}

In the present paper we have applied likelihood methods to the
specific WIMP scenario of proton--philic spin--dependent inelastic
Dark Matter (pSIDM), summarized in Section \ref{sec:spin_idm_scenario}
and introduced in\cite{noi_idm_spin} to explain the DAMA yearly
modulation effect \cite{dama_last} in compliance with the constraints
from other direct detection DM searches. To this am, we obtained
updated ranges for the model parameters $m_{\chi}$ (WIMP mass),
$\delta$ (mass splitting) and $r=c^n/c^p$ (neutron-to-proton coupling
ratio) both in a halo--independent approach and adopting a truncated
Maxwellian for the WIMP velocity distribution, constructing
approximate frequentist confidence intervals from an effective
chi--square including, among others, the latest experimental
constraints from XENON1T\cite{xenon_1t}, PANDAX-II\cite{panda_2017},
SuperCDMS\cite{supercdms_2017} and CDMSlite \cite{cdmslite2_2017}.  In
the halo--independent analysis we have implemented the dependence on
the WIMP velocity distribution through the step--wise parameterization
(\ref{eq:delta_eta01_piecewise}) of the halo-functions
$\tilde{\eta}_0$ and $\tilde{\eta}_1$ defined in Eqs. (\ref{eq:s0})
and (\ref{eq:s1}) for the time--averaged and the time--modulated parts
of the signal. In the pSIDM the WIMP incoming velocities required to
explain the DAMA effect fall in a narrow range close to the escape
velocity so that, in practice, a limited number of steps ($N\le$ 5) in
the the step--wise parameterization of the halo functions was
sufficient to determine the profile likelihood of the parameters.
Specifically, we have allowed the $\tilde{\eta}_{1,k}$'s step values
of the modulated halo function to vary freely, and parameterized the
corresponding $\tilde{\eta}_{0,k}$'s through
Eqs.(\ref{eq:minimal_eta0}) and (\ref{eq:eta_0_min}), i.e. in terms of
the minimal time-averaged halo function compatible to a WIMP
explanation of the DAMA effect. For the calculation of the profile
likelihood of the pSIDM parameters we used emcee\cite{emcee}, a Markov
Chain Montecarlo (MCMC) generator. The numerical procedure required a
large number of evaluations of the expected rates, a time--consuming
task when calculating the relevant experimental response functions at
run time. For this reason in Section \ref{sec:response_functions} we
have introduced expected rates expressions rearranged in terms of
differences of singled--valued integrated response functions suitable
for a fast evaluation through tabulation and interpolation. Examples
of such integrated response functions are provided in
Fig.\ref{fig:response_functions}.

Our frequentist analysis confirms the present viability of the pSIDM
scenario as a possible explanation of the DAMA effect, with the
1--sigma parameter ranges of Eq.(\ref{eq:1sigma_int_halo_indep}) for
the halo--independent case, and of Eq.(\ref{eq:1sigma_int_maxwellian})
when the WIMP velocity distribution is given by a standard truncated
Maxellian (the latter analysis also yields a range for the
WIMP--proton cross section, as illustrated in
Fig.\ref{fig:mchi_sigma_maxwellian}).

Although in the pSIDM scenario the response of neutron--odd targets is
suppressed, our analysis signals some residual tension between DAMA
and $^{129}Xe$ and $^{131}Xe$ in XENON1T and PANDA, since the
best--fit value of the $r=c^n/c^p$ parameter must be tuned to
$r=c^n/c^p\simeq$=-0.03, which corresponds to a cancellation in the
xenon spin--dependent nuclear form factors. In particular, such value
is $\simeq$ 2.4 $\sigma$ away from zero in the halo--independent case
and $\simeq$ 4.8 $\sigma$ away from zero for the Maxwellian case. As
shown in Fig. \ref{fig:rate_xenon_1t}, an order of magnitude
improvement in the exposure of Ref.\cite{xenon_1t} (corresponding to a
full year of data taking) should allow XENON1T to reach the level of
sensitivity required to start probing the pSIDM scenario.

\acknowledgments This research was supported by the Basic Science
Research Program through the National Research Foundation of
Korea(NRF) funded by the Ministry of Education, grant number
2016R1D1A1A09917964.
\appendix

\section{Experimental constraints}
\label{app:exp}
In the present analysis we include the experimental data from
DAMA\cite{dama_last},XENON1T\cite{xenon_1t},
PANDAX-II\cite{panda_2017},  SuperCDMS\cite{supercdms_2017}, CDMSlite
\cite{cdmslite2_2017}, PICO-2L\cite{pico2l_2016}, COUPP\cite{coupp}
and PICASSO\cite{picasso}.

\subsection{DAMA}
We take the DAMA modulation amplitudes normalized to
kg$^{-1}$day$^{-1}$keVee$^{-1}$ from Ref.\cite{dama_last}, assuming a
constant quenching factor $q$=0.3 for sodium and a Gaussian energy
resolution ${\cal
  G}(E^{\prime},E_{ee})=Gauss(E^{\prime}|E_{ee},\sigma)=1/(\sqrt{2\pi}\sigma)exp(-(E^{\prime}-E_{ee})/2\sigma^2)$
with $\sigma$ = 0.0091 (E$_{ee}$/keVee) + 0.448 $\sqrt{E_{ee}}$/keVee
in keV.

\subsection{XENON1T and PANDAX-II}

For xenon detectors the response functions given in
Eq.(\ref{eq:response_function_final}) remain the same with the
expected primary scintillation signal $<S_1>$ in PE (photo-electrons)
in place of the electron--equivalent energy $E_{ee}$ and the quenching
factor $E_{ee}/E_R$ substituted by $<S_1>/E_R$=$g_1 L_y$, with $g_1$
the light collection efficiency and $L_y$ the light yield.

For XENON1T we have assumed zero WIMP candidate events in the range 3
PE$\le S_1 \le$ 30 PE in the lower half of the signal band, as shown
in figure 2 of Ref.\cite{xenon_1t} for the primary scintillation
signal S1 (directly in Photo Electrons, PE) for an exposure of 34.2
days and a fiducial volume of 1042 kg of Xenon. We have used the
efficiency taken from Fig. 1 of \cite{xenon_1t}, a light collection
efficiency $g_1$=0.144, while for the light yield $L_y$ we have used
the NEST model of Ref. \cite{nest} with an electric field E=120 v/cm
and the parameters of Table 1 with the exception of the Lindhard
parameter k=0.15, to reproduce the combined energy curves of Fig. 2b of
\cite{xenon_1t}.

On the other hand for PANDAX-II we included the result of Run 10
\cite{panda_2017} with zero WIMP candidate events in the range 3
PE$\le S_1 \le$ 45 PE in the lower half of the signal band, as shown
in figure 4, for an exposure of 77.1 days and a fiducial mass of 361.5
kg. From the supplemental material provided in \cite{panda_2017} we
have taken the efficiency in Fig.16 , $g_1$=0.1114 and $L_y$ in
Fig.13b.

For both XENON1T and PANDAX-II we have modeled the energy resolution
combining a Poisson fluctuation of the observed primary signal $S_1$
compared to $<S_1>$ and a Gaussian response of the photomultiplier
with $\sigma_{PMT}=0.5$, so that:

\begin{equation}
{\cal G}_{Xe}(E_R,S)=\sum_{n=1}^{\infty}
Gauss(S|n,\sqrt{n}\sigma_{PMT})Poiss(n,<S(E_R)>),
\label{eq:g_xe}  
\end{equation}

\noindent with $Poiss(n,\lambda)=\lambda^n/n!exp(-\lambda)$.

\subsection{PICO-2L, COUPP and PICASSO}

Bubble chambers are threshold experiments, and in the response
functions of Eq.(\ref{eq:response_function_final}) one has $q(E_R)$=1
and:

\begin{equation}
{\cal G}_T(E^{\prime},E_R)={\cal P}_T(E_R)\delta(E^{\prime}-E_R).
  \end{equation}
\noindent For the nucleation probability of Fluorine we take:

\begin{equation}
{\cal P}_F(E_R)=1-\exp\left [-\alpha_T\frac{E_R-E_{th}}{E_{th}} \right ]
\label{eq:nucleation_probability}
\end{equation}

\noindent with $\alpha$=5, while for Iodine we take ${\cal P}_I$=1.

PICO-2L uses $C_3F_8$. Its latest analysis \cite{pico2l_2016} was
performed after an upgrade of the detector that significantly reduced
the background compared to \cite{pico2l}. Only the threshold
$E_{th}$=3.3 keV was analyzed, with a total exposure of 129.0 kg day
and 1 event detected. We use for Fluorine and Carbon the nucleation
probabilities of Fig. 4 of \cite{pico2l}.

COUPP uses a $CF_3I$ target.  For each operating threshold used in
COUPP the corresponding exposure and number of measured events are
summarized in Table \ref{table:coupp}. For fluorine we use the
nucleation probability of Eq.(\ref{eq:nucleation_probability}) with
$\alpha$=0.15.

\begin{table}[t]
\begin{center}
{\begin{tabular}{@{}|c|c|c|c|@{}}
\hline
$E_{th}$ (keV) & exposure (kg day) & measured events  \\
\hline
7.8 & 55.8  & 2 \\
11 & 70   & 3 \\
15.5 & 311.7  & 8 \\
\hline
\end{tabular}}
\caption{For each operating threshold used in COUPP we provide the
  corresponding exposure and number of measured events used in our
  analysis. \label{table:coupp}}
\end{center}
\end{table}

Finally, the PICASSO experiment\cite{picasso} is a bubble
chamber using $C_3 F_8$, operated with eight energy thresholds. For
each of the latter we provide the corresponding numbers of observed
events and statistical fluctuations (normalized to events/kg/day and
used in Eq.(\ref{eq:likelihood_picasso})) in Table
\ref{table:picasso} (extracted from Fig. 5 of Ref.\cite{picasso}). We
use the nucleation probability of Eq.(\ref{eq:nucleation_probability})
with $\alpha_C$=$\alpha_F$=5.

\begin{table}[t]
\begin{center}
{\begin{tabular}{@{}|c|c|c|c|c|@{}}
\hline
$E_{th}$ (keV) & Event rate (events/kg/day) & Fluctuation \\
\hline
1.7   &  -6.0  & 7.1    \\
2.9   &  -0.3  & 1.8    \\
4.1   &  1.6   & 9.0    \\
5.8   &  -0.2  & 9.2    \\
6.9   &  0.0   & 1.3    \\
16.3  &  1.4   & 1.7     \\
 39   &  0.2   & 1.7     \\ 
 55   &  1.3   & 4.7     \\
\hline
\end{tabular}}
\caption{Observed number of events and 1--sigma statistical
  fluctuations (extracted from Fig. 5 of
  Ref.\cite{picasso}) for each operating threshold used in PICASSO.
  \label{table:picasso}}
\end{center}
\end{table}

\subsection {SuperCDMS and CDMSlite}

CDMS operates thermal bolometers for which $q$=1.  The latest
SuperCDMS analysys \cite{supercdms_2017} observes 1 event between 4
and 100 keVnr with an exposure of 1690 kg day. We take the efficiency
from Fig.1 of \cite{supercdms_2017} and $\sigma=\sqrt{0.293^2+0.056^2
  E_{ee}}$.

For CDMSlite we consider the energy bin 0.17 keV$<E^{\prime}<$ 1.1 keV
with a measured count rate of 1.1$\pm$0.2 [keV kg day]$^{-1}$ (Full
Run 2 rate, Table II of Ref. \cite{cdmslite2_2017}). We take the
efficiency from Fig. 4 of \cite{cdmslite2_2017} and
$\sigma=\sqrt{\sigma_E^2+B E_R+(A E_R)^2}$, with $\sigma_E$=9.26 eV,
$A$=5.68$\times 10^{-3}$ and $B$=0.64 eV from Section IV.A.


\begin{thebibliography}{99}

\bibitem{dama_last}
R.~Bernabei et~al., {\it {Final model independent result of
  DAMA/LIBRA-phase1}},  {\em Eur. Phys. J.} {\bf C73} (2013) 2648,
  [\href{http://arxiv.org/abs/1308.5109}{{\tt arXiv:1308.5109}}].

\bibitem{xenon_1t}
{\bf XENON} Collaboration, E.~Aprile et~al., {\it {First Dark Matter Search
  Results from the XENON1T Experiment}},  {\em Phys. Rev. Lett.} {\bf 119}
  (2017), no.~18 181301, [\href{http://arxiv.org/abs/1705.06655}{{\tt
  arXiv:1705.06655}}].

\bibitem{panda_2017}
{\bf PandaX-II} Collaboration, X.~Cui et~al., {\it {Dark Matter Results From
  54-Ton-Day Exposure of PandaX-II Experiment}},  {\em Phys. Rev. Lett.} {\bf
  119} (2017), no.~18 181302, [\href{http://arxiv.org/abs/1708.06917}{{\tt
  arXiv:1708.06917}}].

\bibitem{supercdms_2017}
{\bf SuperCDMS} Collaboration, R.~Agnese et~al., {\it {Results from the Super
  Cryogenic Dark Matter Search (SuperCDMS) experiment at Soudan}},
  \href{http://arxiv.org/abs/1708.08869}{{\tt arXiv:1708.08869}}.

\bibitem{cdmslite2_2017}
{\bf SuperCDMS} Collaboration, R.~Agnese et~al., {\it {Low-Mass Dark Matter
  Search with CDMSlite}},  {\em Submitted to: Phys. Rev. D} (2017)
  [\href{http://arxiv.org/abs/1707.01632}{{\tt arXiv:1707.01632}}].

\bibitem{pico2l_2016}
{\bf PICO} Collaboration, C.~Amole et~al., {\it {Improved dark matter search
  results from PICO-2L Run 2}},  {\em Phys. Rev.} {\bf D93} (2016), no.~6
  061101, [\href{http://arxiv.org/abs/1601.03729}{{\tt arXiv:1601.03729}}].

\bibitem{coupp}
{\bf COUPP} Collaboration, E.~Behnke et~al., {\it {First Dark Matter Search
  Results from a 4-kg CF$_3$I Bubble Chamber Operated in a Deep Underground
  Site}},  {\em Phys. Rev.} {\bf D86} (2012), no.~5 052001,
  [\href{http://arxiv.org/abs/1204.3094}{{\tt arXiv:1204.3094}}]. [Erratum:
  Phys. Rev.D90,no.7,079902(2014)].

\bibitem{haxton1}
A.~L. Fitzpatrick, W.~Haxton, E.~Katz, N.~Lubbers, and Y.~Xu, {\it {The
  Effective Field Theory of Dark Matter Direct Detection}},  {\em JCAP} {\bf
  1302} (2013) 004, [\href{http://arxiv.org/abs/1203.3542}{{\tt
  arXiv:1203.3542}}].

\bibitem{haxton2}
N.~Anand, A.~L. Fitzpatrick, and W.~C. Haxton, {\it {Weakly interacting massive
  particle-nucleus elastic scattering response}},  {\em Phys. Rev.} {\bf C89}
  (2014), no.~6 065501, [\href{http://arxiv.org/abs/1308.6288}{{\tt
  arXiv:1308.6288}}].

\bibitem{inelastic}
D.~Tucker-Smith and N.~Weiner, {\it {Inelastic dark matter}},  {\em Phys. Rev.}
  {\bf D64} (2001) 043502, [\href{http://arxiv.org/abs/hep-ph/0101138}{{\tt
  hep-ph/0101138}}].

\bibitem{factorization}
P.~J. Fox, J.~Liu, and N.~Weiner, {\it {Integrating Out Astrophysical
  Uncertainties}},  {\em Phys. Rev.} {\bf D83} (2011) 103514,
  [\href{http://arxiv.org/abs/1011.1915}{{\tt arXiv:1011.1915}}].

\bibitem{Feldstein_2014}
B.~Feldstein and F.~Kahlhoefer, {\it {A new halo-independent approach to dark
  matter direct detection analysis}},  {\em JCAP} {\bf 1408} (2014) 065,
  [\href{http://arxiv.org/abs/1403.4606}{{\tt arXiv:1403.4606}}].

\bibitem{Fox_2014}
P.~J. Fox, Y.~Kahn, and M.~McCullough, {\it {Taking Halo-Independent Dark
  Matter Methods Out of the Bin}},  {\em JCAP} {\bf 1410} (2014), no.~10 076,
  [\href{http://arxiv.org/abs/1403.6830}{{\tt arXiv:1403.6830}}].

\bibitem{quantifying_feldstein}
B.~Feldstein and F.~Kahlhoefer, {\it {Quantifying (dis)agreement between direct
  detection experiments in a halo-independent way}},  {\em JCAP} {\bf 1412}
  (2014), no.~12 052, [\href{http://arxiv.org/abs/1409.5446}{{\tt
  arXiv:1409.5446}}].

\bibitem{gondolo_out_of_the_bin}
G.~B. Gelmini, A.~Georgescu, P.~Gondolo, and J.-H. Huh, {\it {Extended Maximum
  Likelihood Halo-independent Analysis of Dark Matter Direct Detection Data}},
  {\em JCAP} {\bf 1511} (2015), no.~11 038,
  [\href{http://arxiv.org/abs/1507.03902}{{\tt arXiv:1507.03902}}].

\bibitem{gelmini_assessing_compatibility}
G.~B. Gelmini, J.-H. Huh, and S.~J. Witte, {\it {Assessing Compatibility of
  Direct Detection Data: Halo-Independent Global Likelihood Analyses}},  {\em
  JCAP} {\bf 1610} (2016), no.~10 029,
  [\href{http://arxiv.org/abs/1607.02445}{{\tt arXiv:1607.02445}}].

\bibitem{gelmini_convex_hulls}
G.~B. Gelmini, J.-H. Huh, and S.~J. Witte, {\it {Unified Halo-Independent
  Formalism Derived From Convex Hulls}},
  \href{http://arxiv.org/abs/1707.07019}{{\tt arXiv:1707.07019}}.

\bibitem{scopel_gondolo_unmodulated}
P.~Gondolo and S.~Scopel, {\it {Halo-independent determination of the
  unmodulated WIMP signal in DAMA: the isotropic case}},
  \href{http://arxiv.org/abs/1703.08942}{{\tt arXiv:1703.08942}}.

\bibitem{noi_idm_spin}
S.~Scopel and K.-H. Yoon, {\it {Inelastic dark matter with spin-dependent
  couplings to protons and large modulation fractions in DAMA}},  {\em JCAP}
  {\bf 1602} (2016), no.~02 050, [\href{http://arxiv.org/abs/1512.00593}{{\tt
  arXiv:1512.00593}}].

\bibitem{emcee}
D.~Foreman-Mackey, D.~W. Hogg, D.~Lang, and J.~Goodman, {\it {emcee: The MCMC
  Hammer}},  {\em Publ. Astron. Soc. Pac.} {\bf 125} (2013) 306--312,
  [\href{http://arxiv.org/abs/1202.3665}{{\tt arXiv:1202.3665}}].

\bibitem{lux}
{\bf LUX} Collaboration, D.~S. Akerib et~al., {\it {First results from the LUX
  dark matter experiment at the Sanford Underground Research Facility}},  {\em
  Phys. Rev. Lett.} {\bf 112} (2014) 091303,
  [\href{http://arxiv.org/abs/1310.8214}{{\tt arXiv:1310.8214}}].

\bibitem{lux_2015_reanalysis}
{\bf LUX} Collaboration, D.~S. Akerib et~al., {\it {Improved Limits on
  Scattering of Weakly Interacting Massive Particles from Reanalysis of 2013
  LUX Data}},  {\em Phys. Rev. Lett.} {\bf 116} (2016), no.~16 161301,
  [\href{http://arxiv.org/abs/1512.03506}{{\tt arXiv:1512.03506}}].

\bibitem{lux_complete}
D.~S. Akerib et~al., {\it {Results from a search for dark matter in the
  complete LUX exposure}},  \href{http://arxiv.org/abs/1608.07648}{{\tt
  arXiv:1608.07648}}.

\bibitem{panda_run_8}
{\bf PandaX} Collaboration, A.~Tan et~al., {\it {Dark Matter Search Results
  from the Commissioning Run of PandaX-II}},  {\em Phys. Rev.} {\bf D93}
  (2016), no.~12 122009, [\href{http://arxiv.org/abs/1602.06563}{{\tt
  arXiv:1602.06563}}].

\bibitem{panda_run_9}
{\bf PandaX-II} Collaboration, A.~Tan et~al., {\it {Dark Matter Results from
  First 98.7 Days of Data from the PandaX-II Experiment}},  {\em Phys. Rev.
  Lett.} {\bf 117} (2016), no.~12 121303,
  [\href{http://arxiv.org/abs/1607.07400}{{\tt arXiv:1607.07400}}].

\bibitem{cdms_ge}
{\bf CDMS-II} Collaboration, Z.~Ahmed et~al., {\it {Results from a Low-Energy
  Analysis of the CDMS II Germanium Data}},  {\em Phys. Rev. Lett.} {\bf 106}
  (2011) 131302, [\href{http://arxiv.org/abs/1011.2482}{{\tt
  arXiv:1011.2482}}].

\bibitem{cdms_lite}
{\bf SuperCDMS} Collaboration, R.~Agnese et~al., {\it {Search for Low-Mass
  Weakly Interacting Massive Particles Using Voltage-Assisted Calorimetric
  Ionization Detection in the SuperCDMS Experiment}},  {\em Phys. Rev. Lett.}
  {\bf 112} (2014), no.~4 041302, [\href{http://arxiv.org/abs/1309.3259}{{\tt
  arXiv:1309.3259}}].

\bibitem{super_cdms}
{\bf SuperCDMS} Collaboration, R.~Agnese et~al., {\it {Search for Low-Mass
  Weakly Interacting Massive Particles with SuperCDMS}},  {\em Phys. Rev.
  Lett.} {\bf 112} (2014), no.~24 241302,
  [\href{http://arxiv.org/abs/1402.7137}{{\tt arXiv:1402.7137}}].

\bibitem{cdms_2015}
{\bf SuperCDMS} Collaboration, R.~Agnese et~al., {\it {Improved WIMP-search
  reach of the CDMS II germanium data}},  {\em Phys. Rev.} {\bf D92} (2015),
  no.~7 072003, [\href{http://arxiv.org/abs/1504.05871}{{\tt
  arXiv:1504.05871}}].

\bibitem{kims}
S.~C. Kim et~al., {\it {New Limits on Interactions between Weakly Interacting
  Massive Particles and Nucleons Obtained with CsI(Tl) Crystal Detectors}},
  {\em Phys. Rev. Lett.} {\bf 108} (2012) 181301,
  [\href{http://arxiv.org/abs/1204.2646}{{\tt arXiv:1204.2646}}].

\bibitem{kims_modulation}
Y.~Kim, {\it {Recent progress in KIMS experiment}},  {\em talk given at
  13$^{th}$ International Conference on Topics in Astroparticle and Underground
  Physics, September 8--13 2013, Asilomar, California USA (TAUP2013)}.

\bibitem{kims2}
H.~S. Lee et~al., {\it {Search for Low-Mass Dark Matter with CsI(Tl) Crystal
  Detectors}},  {\em Phys. Rev.} {\bf D90} (2014), no.~5 052006,
  [\href{http://arxiv.org/abs/1404.3443}{{\tt arXiv:1404.3443}}].

\bibitem{spin_n_suppression}
P.~Ullio, M.~Kamionkowski, and P.~Vogel, {\it {Spin dependent WIMPs in DAMA?}},
   {\em JHEP} {\bf 07} (2001) 044,
  [\href{http://arxiv.org/abs/hep-ph/0010036}{{\tt hep-ph/0010036}}].

\bibitem{spin_gelmini}
E.~Del~Nobile, G.~B. Gelmini, A.~Georgescu, and J.-H. Huh, {\it {Reevaluation
  of spin-dependent WIMP-proton interactions as an explanation of the DAMA
  data}},  {\em JCAP} {\bf 1508} (2015), no.~08 046,
  [\href{http://arxiv.org/abs/1502.07682}{{\tt arXiv:1502.07682}}].

\bibitem{simple}
M.~Felizardo et~al., {\it {Final Analysis and Results of the Phase II SIMPLE
  Dark Matter Search}},  {\em Phys. Rev. Lett.} {\bf 108} (2012) 201302,
  [\href{http://arxiv.org/abs/1106.3014}{{\tt arXiv:1106.3014}}].

\bibitem{picasso}
{\bf PICASSO} Collaboration, S.~Archambault et~al., {\it {Constraints on
  Low-Mass WIMP Interactions on $^{19}F$ from PICASSO}},  {\em Phys. Lett.}
  {\bf B711} (2012) 153--161, [\href{http://arxiv.org/abs/1202.1240}{{\tt
  arXiv:1202.1240}}].

\bibitem{pico2l}
{\bf PICO} Collaboration, C.~Amole et~al., {\it {Dark Matter Search Results
  from the PICO-2L C$_3$F$_8$ Bubble Chamber}},  {\em Phys. Rev. Lett.} {\bf
  114} (2015), no.~23 231302, [\href{http://arxiv.org/abs/1503.00008}{{\tt
  arXiv:1503.00008}}].

\bibitem{pico60}
{\bf PICO} Collaboration, C.~Amole et~al., {\it {Dark Matter Search Results
  from the PICO-60 CF$_3$I Bubble Chamber}},  {\em Submitted to: Phys. Rev. D}
  (2015) [\href{http://arxiv.org/abs/1510.07754}{{\tt arXiv:1510.07754}}].

\bibitem{noi_eft_spin}
S.~Scopel, K.-H. Yoon, and J.-H. Yoon, {\it {Generalized spin-dependent
  WIMP-nucleus interactions and the DAMA modulation effect}},  {\em JCAP} {\bf
  1507} (2015), no.~07 041, [\href{http://arxiv.org/abs/1505.01926}{{\tt
  arXiv:1505.01926}}].

\bibitem{gondolo_generalized}
E.~Del~Nobile, G.~Gelmini, P.~Gondolo, and J.-H. Huh, {\it {Generalized Halo
  Independent Comparison of Direct Dark Matter Detection Data}},  {\em JCAP}
  {\bf 1310} (2013) 048, [\href{http://arxiv.org/abs/1306.5273}{{\tt
  arXiv:1306.5273}}].

\bibitem{gondolo_scopel}
P.~Gondolo and S.~Scopel, {\it {Halo-independent determination of the
  unmodulated WIMP signal in DAMA: the isotropic case}},  {\em JCAP} {\bf 1709}
  (2017), no.~09 032, [\href{http://arxiv.org/abs/1703.08942}{{\tt
  arXiv:1703.08942}}].

\bibitem{Ibarra_Rappelt_2017}
A.~Ibarra and A.~Rappelt, {\it {Optimized velocity distributions for direct
  dark matter detection}},  {\em JCAP} {\bf 1708} (2017), no.~08 039,
  [\href{http://arxiv.org/abs/1703.09168}{{\tt arXiv:1703.09168}}].

\bibitem{profile_likelihood}
F.~Feroz, K.~Cranmer, M.~Hobson, R.~Ruiz~de Austri, and R.~Trotta, {\it
  {Challenges of Profile Likelihood Evaluation in Multi-Dimensional SUSY
  Scans}},  {\em JHEP} {\bf 06} (2011) 042,
  [\href{http://arxiv.org/abs/1101.3296}{{\tt arXiv:1101.3296}}].

\bibitem{vesc_2014}
T.~Piffl et~al., {\it {The RAVE survey: the Galactic escape speed and the mass
  of the Milky Way}},  {\em Astron. Astrophys.} {\bf 562} (2014) A91,
  [\href{http://arxiv.org/abs/1309.4293}{{\tt arXiv:1309.4293}}].

\bibitem{v0_koposov}
S.~E. Koposov, H.-W. Rix, and D.~W. Hogg, {\it Constraining the milky way
  potential with a six-dimensional phase-space map of the gd-1 stellar stream},
   {\em The Astrophysical Journal} {\bf 712} (2010), no.~1 260.

\bibitem{spin_form_factors}
M.~T. Ressell and D.~J. Dean, {\it {Spin dependent neutralino - nucleus
  scattering for A approximately 127 nuclei}},  {\em Phys. Rev.} {\bf C56}
  (1997) 535--546, [\href{http://arxiv.org/abs/hep-ph/9702290}{{\tt
  hep-ph/9702290}}].

\bibitem{nest}
B.~Lenardo, K.~Kazkaz, A.~Manalaysay, J.~Mock, M.~Szydagis, and M.~Tripathi,
  {\it {A Global Analysis of Light and Charge Yields in Liquid Xenon}},  {\em
  IEEE Trans. Nucl. Sci.} {\bf 62} (2015), no.~6 3387--3396,
  [\href{http://arxiv.org/abs/1412.4417}{{\tt arXiv:1412.4417}}].

\end{thebibliography}
\end{document}